\documentclass[11pt]{article}
\usepackage{times}
\usepackage{latexsym}
\usepackage{graphicx}
\usepackage{url}
\usepackage{amsmath}
\usepackage{amsthm}
\usepackage{amssymb}

\newcommand{\be}{\begin{equation}}
\newcommand{\ee}{\end{equation}}
\newcommand{\ba}{\begin{array}}
\newcommand{\ea}{\end{array}}
\newcommand{\bea}{\begin{eqnarray}}
\newcommand{\eea}{\end{eqnarray}}

\newcommand{\calM}{{\cal M }}

\newcommand{\calT}{{\cal T }}
\newcommand{\calZ}{{\cal Z }}

\newcommand{\calG}{{\cal G }}
\newcommand{\calX}{{\cal X }}
\newcommand{\ZZ}{\mathbb{Z}}
\newcommand{\CC}{\mathbb{C}}

\newcommand{\nn}{\nonumber}

\newcommand{\pf}{\mathop{\mathrm{Pf}}\nolimits}
\newcommand{\eqdef}{\stackrel{\scriptscriptstyle \mathrm{def}}{=}}
\newcommand{\perfectm}[2]{\mathrm{PerfMatch}(#1,#2)}

\newtheorem{dfn}{Definition}
\newtheorem{lemma}{Lemma}
\newtheorem{prop}{Proposition}
\newtheorem{theorem}{Theorem}

\newtheorem{cor}{Corollary}

\title{Contraction of matchgate tensor networks on non-planar graphs}
\author{Sergey Bravyi\footnote{
IBM T.J. Watson Research Center,
Yorktown Heights, NY 10598}}

\begin{document}
\maketitle
\begin{abstract}
A tensor network is a product of tensors associated with vertices of some graph $G$ such that every
edge of $G$ represents a  summation (contraction) over a matching pair of indexes.
 It was shown recently by Valiant, Cai, and Choudhary that tensor networks can be efficiently contracted
on planar graphs if components of every tensor obey a system of quadratic equations known as matchgate identities.
Such tensors are referred to as  {\it matchgate tensors}. The present paper provides an alternative approach to
contraction of matchgate tensor networks that easily extends to non-planar graphs.
Specifically, it is shown that a matchgate tensor network on a graph $G$ of genus $g$ with $n$ vertices
can be contracted in time $T=poly(n) + O(m^3)\, 2^{2g}$ where
$m$ is the minimum number of edges one has to remove from $G$ in order to make it planar.
Our approach makes use
of anticommuting (Grassmann) variables and Gaussian integrals.
\end{abstract}

\newpage

\tableofcontents

\newpage

\section{Introduction and summary of results}

Contraction of tensor networks is a computational problem having a variety of applications ranging from
simulation of classical and quantum spin systems~\cite{MarkovShi05,Verstraete04,Vidal05,Vidal07,Levin06} to
computing capacity of data storage devices~\cite{Schwartz07}.
Given the tremendous amount of applications it is important to identify special classes of tensor networks that can be
contracted efficiently.  For example, Markov and Shi  found a linear time algorithm
for contraction of tensor networks on trees and graphs with a bounded treewidth~\cite{MarkovShi05}.
An important class of graphs that do not fall into this category are planar graphs.
Although contraction of an arbitrary tensor network on a planar graph is a hard problem,
it has been known for a long time that the generating function of
perfect matchings known as the {\it matching sum}
can be computed efficiently on planar graphs
for arbitrary (complex) weights using the Fisher-Kasteleyn-Temperley (FKT) method, see~\cite{Fisher61,Kasteleyn61,Temperley61}. It is based on the observation that the matching sum
can be related to Pfaffian of  a weighted adjacency matrix (known as the Tutte matrix).
The FKT method also yields an efficient algorithm for computing
the partition function of spin models reducible to the matching sum, most notably, the Ising model
on a planar graph~\cite{Barahona81}.
Recently the FKT method has been generalized to the matching sum
of non-planar graphs with a bounded genus~\cite{Gallucio99,Zecchina01,Reshetikhin06}.

Computing the matching sum can be regarded as a special
case of a tensor network contraction. It is therefore desirable to characterize
precisely the class of tensor networks
that can be contracted efficiently using the FKT method. This problem has been solved by Valiant~\cite{Valiant02a,Valiant07}
and in the subsequent works by Cai and Choudhary~\cite{Cai2006a,Cai2006b,Cai2006c}.
Unfortunately, it turned out  that the
matching sum of planar graphs essentially provides the most general tensor network in this class,
see~\cite{Cai2006a,Cai2006c}. Following~\cite{Cai2006a}
we shall call such networks {\it matchgate tensor networks}, or simply matchgate networks.
A surprising discovery made in~\cite{Cai2006b} is that matchgate tensors can be characterized  by a simple
system of quadratic equations known as {\it matchgate identities} which does not make references to any graph theoretical
concepts. Specifically, given a tensor $T$ of rank $n$ with complex-valued components
$T(x)=T_{x_1,\, x_2,\ldots,\, x_n}$ labeled by $n$-bit strings $x\in \{0,1\}^n$ one calls $T$ a {\it matchgate tensor}, or simply a matchgate, if
\be
\label{matchgate_identities}
\sum_{a\, : \, x_a\ne y_a} T({x\oplus e^a}) \, T({y\oplus e^a}) \, (-1)^{x_{1} + \ldots + x_{a-1} + y_{1} + \ldots +y_{a-1}} =0
\quad \mbox{for all} \quad x,y\in \{0,1\}^n.
\ee
Here $e^a$ denotes a string in which the $a$-th bit is $1$ and all other bits are $0$. The symbol
$\oplus$ stands for a bit-wise XOR of binary strings.
For example, a simple algebra shows that
a tensor of rank $n=1,2,3$ is a matchgate iff it is either even or odd\footnote{A tensor $T$ is called even  (odd)
if $T(x)=0$ for all strings $x$ with odd (even)  Hamming weight.}. Furthermore, an even  tensor of rank $4$ is a matchgate iff
\be
\label{Grank4}
-T(0000)\, T(1111) + T(1100) \, T(0011) - T(1010)\, T(0101) + T(1001)\, T(0110) = 0.
\ee
A matchgate network is a tensor network in which every tensor is a matchgate.

The purpose of the present paper is two-fold.
Firstly, we develop a formalism that allows one
to perform {\it partial contractions} of matchgate networks, for example, contraction of a single edge
combining its endpoints into a single vertex.
More generally, the formalism allows one to contract any connected planar subgraph $G$ of the network
into a single vertex $u(G)$ by "integrating out" all internal edges of $G$. The number of parameters describing
the contracted tensor assigned to $u(G)$ is independent of the size of $G$. It depends only on the
number of "external" edges connecting $G$ to the rest of the network.
This is the main distinction of
our formalism compared to the original matchgate formalism of Valiant~\cite{Valiant02a}.
The ability to implement partial contractions
may be useful for designing efficient parallel contraction algorithms. More importantly,
we show that it yields a faster contraction algorithm for matchgate networks on non-planar graphs.

Our formalism makes use of  anticommuting (Grassmann) variables such that a tensor of rank $n$ is represented by a
generating function of $n$ Grassmann variables. A matchgate tensor is shown to have a Gaussian generating
function that depends on $O(n^2)$ parameters. The matchgate identities Eq.~(\ref{matchgate_identities})
can be described by a first-order differential equation
making manifest their underlying symmetry.
Contraction of tensors is equivalent to convolution of their generating functions.
Contraction of matchgate tensors can be performed efficiently
using the standard Gaussian integration technique.
We use the formalism to prove that a tensor satisfies matchgate identities if and only if it can be represented by
the matching sum on some planar graph. It reproduces the result obtained earlier by Cai and Choudhary~\cite{Cai2006b,Cai2006c}.
Our approach also reveals that the notion of a matchgate tensor
is equivalent to the one of a Gaussian operator introduced in~\cite{Bravyi05} in the context of quantum computation.

Secondly, we describe an improved algorithm for contraction of matchgate networks on non-planar graphs.
Let $\Sigma$ be a standard oriented closed surface of genus $g$, i.e., a sphere with $g$ handles.
\begin{dfn}
\label{dfn:planar_cut}
Given a graph $G=(V,E)$ embedded into a surface  $\Sigma$ we shall say that $G$ is contractible
if there exists a region $D\subset \Sigma$ with topology of a disk  containing
all vertices and all edges of $G$.
A subset of edges $M\subseteq E$ is called a planar cut of $G$ if
a graph $G_M=(V,E\backslash M)$ is contractible.
\end{dfn}
A {\it contraction value} $c(\calT)$ of a  tensor network $\calT$ is a complex number obtained by
contracting all tensors of $\calT$.
Our main result is as follows.
\begin{theorem}
\label{thm:main}
Let $\calT$ be a matchgate tensor network on a
graph $G=(V,E)$ with $n$ vertices embedded into a surface of genus $g$.
Assume we are given a planar cut of $G$ with $m$ edges.
Then the  contraction value $c(\calT)$ can be computed in time $T=O((n+m)^6)+ O(m^3)\, 2^{2g}$.
If $G$ has a bounded vertex degree, one can compute $c(\calT)$ in time
$T=O((n+m)^3)+ O(m^3)\, 2^{2g}$.
\end{theorem}
If a network has a small planar cut, $m\ll n$,  the theorem provides a speedup
for computing the matching sum and the partition function of the Ising model
compared to the FKT method.
For example, computing the matching sum of a graph $G$ as above
by the FKT method would require time $T=O(n^3)\, 2^{2g}$ since the matching sum
is expressed as a linear combination of $2^{2g}$ Pfaffians where each Pfaffian
involves a matrix of size $n\times n$, see~\cite{Gallucio99,Zecchina01,Reshetikhin06},
and since Pfaffian of an $n\times n$ matrix can be computed in time $O(n^3)$,
see Remark~2 below.
In contrast to the FKT method, our algorithm is divided into two stages.
At the first stage that requires time $O((n+m)^6)$  one performs a partial contraction
of the planar subgraph $G_M$ determined by the given planar cut $M$, see Def.~\ref{dfn:planar_cut}.
 The contraction reduces
the number of edges in a network down to $m$  without changing the genus\footnote{If the initial network
represents a matchings sum, the first stage of the algorithm would require only time $O((n+m)^3)$.}.
The first stage of the algorithm yields a new network $\calT'$ with a single vertex and $m$ self-loops such that
$c(\calT')=c(\calT)$. At the second stage
one contracts the network $\calT'$ by expressing the contraction value $c(\calT')$ as a linear
combination of $2^{2g}$ Pfaffians similar to the FKT method.
However  each Pfaffian involves a matrix  of size only $O(m)\times O(m)$.

\noindent
{\it Remark 1:}
The statement of the theorem assumes that all tensors are specified by their generating functions.
Thus a matchgate tensor of rank $d$ can be specified by $O(d^2)$ parameters, see Section~\ref{sec:matchgates}
for details. The ordering of indexes in any tensor must be consistent with the orientation of a surface.
See Section~\ref{subs:tensor_networks} for a formal definition of tensor networks.

\noindent
{\it Remark 2:}  Recall that Pfaffian of an $n\times n$ antisymmetric matrix $A$ is defined as
\[
\pf{(A)}=\left\{\ba{rcl}
0 &\mbox{if}& \mbox{$n$ is odd},\\
\frac1{2^n\, n!} \sum_{\sigma\in S_n} \mathrm{sgn}(\sigma) \, A_{\sigma(1),\sigma(2)}\,  A_{\sigma(3),\sigma(4)}\cdots A_{\sigma(n-1),\sigma(n)} &\mbox{if} & \mbox{$n$ is even}.\\
\ea\right.
\]
where $S_n$ is the symmetric group and $\mathrm{sgn}(\sigma)=\pm 1$ is the parity of a permutation $\sigma$.
One can efficiently compute Pfaffian  up to a sign using an identity
$\pf{(A)}^2=\det{(A)}$. However, in order to compute a linear combination of several Pfaffians one needs
to know the sign exactly.
One can directly compute $\pf{(A)}$  using the combinatorial
algorithm by Mahajan et al~\cite{Mahajan99} in time $O(n^4)$.
Alternatively, one can use Gaussian elimination to find an invertible matrix $U$ such that
$U^T\, A\, U$ is block-diagonal with all blocks of size $2\times 2$. It requires time $O(n^3)$.
Then $\pf{(A)}$ can be computed using an identity $\pf{(U\, A\, U^T)}=\det{(U)}\, \pf{(A)}$.
This method yields $O(n^3)$ algorithm although it is less computationally stable
compared to the combinatorial algorithm of~\cite{Mahajan99}.

\section{Some definitions and notations}
\subsection{Tensor networks}
\label{subs:tensor_networks}

Throughout this paper a tensor of rank $d$ is a $d$-dimensional complex array $T$
in which the indexes take values $0$ and $1$. Given a binary string of indexes  $x=(x_1 x_2 \ldots x_d)$
we shall denote the corresponding component $T_{x_1 x_2 \ldots x_d}$ as $T(x)$.

A tensor network is a product of tensors whose indexes are pairwise contracted.
More specifically, each tensor is represented by a vertex of some graph
$G=(V,E)$, where  $V$ is a set of vertices and $E$ is a set of edges.
The graph may have self-loops and multiple edges.
 For every edge $e\in E$ one defines a
variable $x(e)$ taking values $0$ and $1$.  A bit string $x$ that assigns a particular value
to every variable $x(e)$ is called an {\it index  string}. A set of all possible index strings will be denoted
$\calX(E)$. In order to define a tensor network on $G$ one has to order edges incident to every
vertex.
We shall assume that $G$  is specified by its {\it incidence list}, i.e., for every vertex $u$
one specifies an ordered list of edges incident to $u$ which will be denoted $E(u)$.
Thus
$E(u)=\{e_1^u,\ldots,e_{d(u)}^u\}$ where $e^u_j\in E$ for all $j$. Here $d(u)=|E(u)|$ is the degree of $u$.
If a vertex $u$ has one or several self-loops, we assume that every self-loop appears in the list $E(u)$
twice (because it will represent contraction of two indexes). For example, a vertex with one self-loop
and no other incident edges has degree $2$.
 A tensor network on  $G$  is a collection of tensors
$\calT=\{T_u\}_{u\in V}$ labeled by vertices of $G$
such that a tensor $T_u$ has rank $d(u)$.
A {\it contraction value} of a network $\calT$ is defined as
\be
c(\calT)=\sum_{x\in \calX(E)} \prod_{u\in V} \,T_u(x(e^u_1)\ldots x(e^u_{d(u)})).
\ee
Thus the contraction value can be computed by taking a tensor product of all tensors $\{T_u\}$
and then contracting those pairs of indexes that correspond to the same edge of the graph.
By definition, $c(\calT)$ is a complex number (tensor of rank $0$).

It will be implicitly assumed
throughout this paper that a tensor network is defined on a graph $G$
embedded into a closed oriented surface $\Sigma$. We require that the order of edges incident to any
vertex $u$ must agree with the order in which the edges appear if one circumnavigates $u$ counterclockwise.
Thus the order on any set $E(u)$ is completely specified by the choice of the first edge $e^u_1\in E(u)$.
If the surface $\Sigma$ has genus $g$ we shall say that $G$ has genus $g$ (it may or may not be the
minimal genus for which the embedding of $G$ into $\Sigma$ is possible).

\subsection{Anticommuting variables}
In this section we introduce notations pertaining to the Grassmann algebra
and anticommuting variables
(see the textbook~\cite{ItzDr} for more details).
Consider a set of formal variables $\theta=(\theta_1,\ldots,\theta_n)$
subject to multiplication rules
\be
\label{grassmann}
\theta_a^2=0, \quad \theta_a \theta_b+\theta_b\theta_a=0 \quad \mbox{for all} \quad a,b.
\ee
The Grassmann algebra $\calG(\theta)$
is the algebra of complex polynomials in variables $\theta_1,\ldots,\theta_n$
factorized over the ideal generated by Eq.~(\ref{grassmann}).
Equivalently, $\calG(\theta)$ is the exterior algebra of the vector space $\CC^n$,
where each variable $\theta_a$ is regarded as a basis vector of $\CC^n$.
More generally, the variables $\theta_a$ may be labeled by elements of an arbitrary finite set $X$
(in our case the variables will be associated with edges or vertices of a graph).
A  linear basis of $\calG(\theta)$ is spanned by $2^n$ monomials in variables $\theta_a$.
Namely, for any subset  $M\subseteq \{1,\ldots,n\}$ define a {\it normally ordered} monomial
\be
\label{normal_order}
\theta(M)=\prod_{a\in M} \theta_a
\ee
where the indexes increase from the left to the right.
If the variables are labeled by elements of some set $X$,
one can define the  normally ordered monomials
$\theta(M)$, $M\subseteq X$ by choosing some order on $X$.
Let us  agree that $\theta(\emptyset)=I$.
Then an arbitrary element $f\in \calG(\theta)$ can be written as
\be
\label{linear_basis}
f=\sum_{M\subseteq \{1,\ldots,n\}} f(M)\, \theta(M), \quad f(M)\in \CC.
\ee
 We shall use notations $f$ and $f(\theta)$
interchangeably meaning that $f$ can be regarded as a function of anticommuting variables
$\theta=(\theta_1,\ldots,\theta_n)$.  Accordingly, elements of the Grassmann algebra will be referred to as
functions.  In particular, $I$ is regarded as a constant function.
A function $f(\theta)$ is called even (odd) if it is a linear combination of monomials $\theta(M)$
 with even (odd)  degree.
 Even functions span the central subalgebra of $\calG(\theta)$.

We shall often consider several species of Grassmann variables, for example, $\theta=(\theta_1,\ldots,\theta_n)$
and $\eta=(\eta_1,\ldots,\eta_k)$.
It is always understood that  different variables anticommute.
For example, a function $f(\theta,\eta)$ must be regarded as an element of the Grassmann
algebra $\calG(\theta,\eta)$, that is,
a linear combination of monomials in  $\theta_1,\ldots,\theta_n$
and $\eta_1,\ldots,\eta_k$.

A partial derivative over a variable $\theta_a$  is a linear map  $\partial_a\, : \, \calG(\theta) \to \calG(\theta)$ defined by
requirement $\partial_a\cdot I=0$ and
the Leibniz rule
\[
\partial_a\cdot (\theta_b \, f) = \delta_{a,b}\, f - \theta_b (\partial_a\cdot f).
\]
More explicitly, given any function $f\in \calG(\theta)$, represent it as $f(\theta)=f_0+\theta_a\, f_1$,
where $f_0, f_1\in \calG(\theta)$ do not depend on $\theta_a$. Then $\partial_a\, f=f_1$.
It follows that $\partial_a\cdot \theta_a=I$, $\partial_a \theta_b=-\theta_b \partial_a$,
 $\partial_a \partial_b=-\partial_b\partial_a$  for $a\ne b$ and $\partial_a^2=0$.

A linear change of variables $\theta_a=\sum_{b=1}^n U_{a,b}\, \tilde{\theta}_b$
with invertible matrix $U$
induces an automorphism of the algebra $\calG(\theta)$ such that $f(\theta)\to f(\tilde{\theta})$. The corresponding
transformation of partial derivatives is
\be
\label{derivative_change}
\partial_a= \sum_{b=1}^n (U^{-1})_{b,a}\,  \tilde{\partial}_b.
\ee

\subsection{Gaussian integrals}
\label{subs:Gintegral}
Let  $\theta=(\theta_1,\ldots,\theta_n)$ be a set of Grassmann variables.
An integral over a variable $\theta_a$ denoted by $\int d\theta_a$ is a linear map
from  $\calG(\theta_1,\ldots,\theta_n)$ to  $\calG(\theta_1,\ldots,\hat{\theta}_a,\ldots,\theta_n)$,
where $\hat{\theta}_a$ means that the variable $\theta_a$ is omitted.
 To define an integral $\int d\theta_a \, f(\theta)$, represent the function $f$ as
$f=f_0+\theta_a\, f_1$, where $f_0,f_1\in \calG(\theta_1,\ldots,\hat{\theta}_a,\ldots,\theta_n)$.
Then $\int d\theta_a f(\theta)=f_1$.  Thus one can compute the integral $\int d\theta_a f(\theta)$
by first computing the derivative $\partial_a\cdot f(\theta)$ and then excluding the variable $\theta_a$
from the list of variables of $f$.

Given an ordered set of Grassmann variables $\theta=(\theta_1,\ldots,\theta_n)$ we shall
use a shorthand notation
\[
\int D\theta = \int d\theta_n \cdots \int d\theta_2 \int d\theta_1.
\]
Thus $\int D\theta$ can be regarded as a linear functional on $\calG(\theta)$,
or as a linear map from $\calG(\theta,\eta)$ to $\calG(\eta)$, and so on.
The action of  $\int D\theta$ on the normally ordered monomials is as follows
 \be
\label{I1}
\int D\theta \, \theta(M) = \left\{ \ba{rcl}
1 &\mbox{if}& M=\{1,2,\ldots,n\},\\
0 && \mbox{otherwise}.\\
\ea\right.
\ee
Similarly, if one regards $\int D\theta$ as a linear map from $\calG(\theta,\eta)$ to $\calG(\eta)$
then
\[
\int D\theta \, \theta(M)\, \eta(K)  = \left\{ \ba{rcl}
\eta(K) &\mbox{if}& M=\{1,2,\ldots,n\},\\
0 && \mbox{otherwise}.\\
\ea\right.
\]
Although this definition assumes that both  variables $\theta$, $\eta$
have a normal ordering, the integral $\int D\theta$ depends only on
the ordering of $\theta$.

One can easily check that integrals over different variables anticommute,
$\int d\theta_a \int d\theta_b = -\int d\theta_b \int d\theta_a$ for $a\ne b$. More generally,
if $\theta=(\theta_1,\ldots,\theta_n)$ and $\eta=(\eta_1,\ldots,\eta_k)$ then
\be
\label{int_com}
\int D\theta \, \int D\eta =(-1)^{nk}\,  \int D\eta \, \int D\theta.
\ee
Under a linear change of variables $\theta_a=\sum_{b=1}^n U_{a,b}\, \eta_b$ the
integral transforms as
\be
\label{int_change}
\int D\theta=\det{(U)}\, \int D\eta.
\ee
In the rest of the section we consider two species of Grassmann variables
 $\theta=(\theta_1,\ldots,\theta_n)$ and $\eta=(\eta_1,\ldots,\eta_k)$.
Given an antisymmetric $n\times n$ matrix $A$ and any $n\times k$ matrix $B$, define quadratic forms
\[
\theta^T\, A\, \theta = \sum_{a,b=1}^n A_{a,b} \, \theta_a\, \theta_b,
\quad
\theta^T\, B\,  \eta=\sum_{a=1}^n\sum_{b=1}^k  B_{a,b}\, \theta_a\, \eta_b.
\]
Gaussian integrals over Grassmann variables are defined as follows.
\be
\label{Gaussian_Integrals}
I(A)\eqdef \int  D\theta \, \exp{\left( \frac12 \, \theta^T\, A \, \theta\right)}
\quad  \mbox{and} \quad
I(A,B)\eqdef \int  D\theta  \, \exp{\left( \frac12 \, \theta^T\, A \, \theta + \theta^T\, B\,  \eta\right)}.
\ee
Thus $I(A)$ is just a complex number while $I(A,B)$ is an element of  $\calG(\eta)$.
Below we present the standard formulas for the Gaussian integrals.
Firstly,
\be
\label{GI1}
I(A)=\pf{(A)}.
\ee
Secondly,  if $A$ is an invertible matrix then
\be
\label{GI2}
I(A,B)=\pf{(A)}\, \exp{\left( \frac12 \, \eta^T\, B^T A^{-1} B\, \eta\right)}.
\ee
Assume now that $A$ has rank
$m$ for some even\footnote{Note that antisymmetric matrices always have even rank.} integer $0\le m\le n$.
Choose any invertible matrix $U$ such that  $AU$ has zero columns $m+1,\ldots,n$.
(This is equivalent to finding a basis of $\CC^n$ such that the last $n-m$ basis vectors belong
to the zero subspace of $A$.)  Then
\[
U^T\, A\, U = \left[ \ba{cc} A_{11} & 0 \\ 0 & 0 \\ \ea \right],
\]
for some invertible $m\times m$ matrix $A_{11}$. Introduce also matrices $B_1$, $B_2$ of size $m\times k$ and
$(n-m)\times k$ respectively such that
\[
U^T\, B = \left[ \ba{c} B_1 \\ B_2 \\ \ea \right].
\]
Performing a change of variables $\theta=U\tilde{\theta}$ in Eq.~(\ref{Gaussian_Integrals}) and introducing
variables $\tau=(\tau_1,\ldots,\tau_m)$ and $\mu=(\mu_1,\ldots,\mu_{n-m})$
such that $\tilde{\theta}=(\tau,\mu)$   one gets
\[
I(A,B)=\det{(U)} \int D\tau \exp{\left( \frac12 \, \tau^T \, A_{11} \, \tau + \tau^T\, B_1\, \eta\right)}
\; \int D\mu \exp{\left( \mu^T\, B_2\, \eta\right)}.
\]
Here we have taken into account Eqs.~(\ref{int_com},\ref{int_change}).
Applying Eq.~(\ref{GI2}) to the first integral one gets
\be
\label{GI3}
I(A,B)=
\pf{(A_{11})}\,\det{(U)}\,
\exp{\left( \frac12 \, \eta^T\, B_1^T (A_{11})^{-1} B_1\, \eta\right)}\;
\int D\mu \exp{\left( \mu^T\, B_2\, \eta\right)}.
\ee
One can easily check that $\int D\mu \exp{\left( \mu^T\, B_2\, \eta\right)}=0$ if
the rank of $B_2$  is smaller than the number of variables in $\mu$, that is, $n-m$.
Since $B_2$ has only $k$ columns we conclude that
\[
I(A,B)=0 \quad  \mbox{unless} \quad  m\ge n-k.
\]
Therefore in the non-trivial case $I(A,B)\ne 0$ the matrices $B_1^T (A_{11})^{-1} B_1$ and $B_2$ specifying $I(A,B)$
have size $k\times k$ and $k'\times k$ for some $k'\le k$.
It means that $I(A,B)$ can be specified by $O(k^2)$ bits. One can compute
$I(A,B)$  in time $O(n^3+n^2 k)$. Indeed,
one can use Gaussian elimination to find $U$,
compute $\det{(U)}$ and $\pf{(A_{11})}$ in time $O(n^3)$. The  matrix $A_{1,1}^{-1}$
can be computed in time $O(n^3)$. Computing the matrices $B_1,B_2$ requires time $O(n^2 k)$.

The formula Eq.~(\ref{GI3}) will be our main tool for contraction of matchgate tensor networks.

\section{Matchgate tensors}
\label{sec:matchgates}

\subsection{Basic properties of matchgate tensors}
Although the definition of a matchgate tensor in terms of the matchgate identities
Eq.~(\ref{matchgate_identities}) is very simple, it is neither very insightful nor
very useful. Two equivalent but more operational definitions will be given in Sections~\ref{subs:matchgate=gaussian},
\ref{subs:matchgate=matchsum}.
Here we list  some basic properties of matchgate tensors that can be derived directly from Eq.~(\ref{matchgate_identities}).
In particular, following the approach of~\cite{Cai2006b}, we prove that a matchgate tensor of rank $n$
can be specified by a {\it mean vector} $z\in \{0,1\}^n$ and a {\it covariance matrix} $A$
of size $n\times n$.
\begin{prop}
Let $T$ be a matchgate tensor of rank $n$.
For any $z\in \{0,1\}^n$ a tensor $T'$ with components $T'(x)=T(x\oplus z)$ is a matchgate tensor.
\end{prop}
\begin{proof}
Indeed, make a change of variables $x\to x\oplus z$, $y\to y\oplus z$ in the  matchgate identities
\end{proof}
Let $T$ be a non-zero matchgate tensor of rank $n$.
Choose any  string $z$ such that $T(z)\ne 0$ and define a new tensor $T'$ with components
\[
T'(x)=\frac{T(x\oplus z)}{T(z)}, \quad x\in \{0,1\}^n,
\]
such that $T'$ is a matchgate and $T'(0^n)=1$. Introduce an antisymmetric $n\times n$ matrix $A$ such that
\[
A_{a,b}=\left\{ \ba{rcl}
T'(e^a\oplus a^b) &\mbox{if}&  a<b, \\
-T'(e^a\oplus a^b) &\mbox{if} &  a>b,\\
0 &\mbox{if} & a=b.\\
\ea
\right.
\]
\begin{prop}
\label{prop:zA}
For any $x\in \{0,1\}^n$
\[
T'(x)=\left\{ \ba{ccl}
\pf{(A(x))} &\mbox{if} & \mbox{$x$ has even weight}\\
0 &\mbox{if} & \mbox{$x$ has odd weight}\\
\ea\right.,
\]
where $A(x)$ is a matrix obtained from $A$ by removing all rows and columns $a$ such that $x_a=0$.
\end{prop}
\begin{proof}
Let us prove the proposition by induction in the weight of $x$.
Choosing $x=0^n$ and $y=e^a$ in the matchgate identities Eq.~(\ref{matchgate_identities})
one gets $T'(e^a)=0$ for all $a$. Similarly, choosing $x=e^b$ and $y=e^a$ with $a< b$
one gets $T'(e^a\oplus e^b)=A_{a,b}=\pf{(A(e^a \oplus  e^b))}$.
Thus the proposition is true for $|x|=1,2$. Assume it is true for all strings $x$ of weight $\le k$.
For any string $x$ of weight $k+1$ and any $a$ such that $x_a=0$ apply the matchgate identities Eq.~(\ref{matchgate_identities})
with $x$ and $y=e^a$. After simple algebra one gets
\[
T'(x\oplus e^a)=\sum_{b\, :\, x_b=1} A_{a,b}\, T'(x\oplus e^b) \, (-1)^{\eta(a,b)}, \quad \eta(a,b)=\sum_{j=a}^{b-1} x_j.
\]
Noting that $x\oplus e^b$ has weight $k$ and applying the induction hypothesis one gets
\[
T'(x\oplus e^a)=\sum_{b\, :\, x_b=1} A_{a,b}\, \pf{(A(x\oplus e^b))} \, (-1)^{\eta(a,b)}
\]
for even $k$ and $T'(x\oplus e^a)=0$ for odd $k$.  Thus $T'(y)=0$ for all odd strings of weight $k+2$.
Furthermore,
let non-zero bits of $x\oplus e^b$ be located at positions $j_1<j_2<\ldots<j_k$.
Note that the sign of $A_{a,b}\, (-1)^{\eta(a,b)}$ coincides with the parity of
a permutation that orders elements in a set $[a,b,j_1,j_2,\ldots,j_k]$. Therefore, by definition of
Pfaffian one gets $T'(x\oplus e^a)=\pf{(A(x\oplus e^a)})$.
\end{proof}
Thus one can regard the vector $z$ and the matrix $A$ above as analogues of a mean vector
and a covariance matrix for Gaussian states of fermionic modes, see for instance~\cite{Bravyi05}.
Although Proposition~\ref{prop:zA} provides a concise description of a matchgate tensor, it is not
very convenient for contracting matchgate networks because the mean vector $z$ and the covariance
matrix $A$ are not uniquely defined.
\begin{cor}
\label{cor:parity}
Any matchgate tensor is either even or odd.
\end{cor}
\begin{proof}
Indeed, the proposition above implies that if a matchgate tensor $T$ has even (odd) mean vector it is an even (odd) tensor.
\end{proof}

\subsection{Describing a tensor by a generating function}
\label{subs:generating}
Let $\theta=(\theta_1,\ldots,\theta_n)$ be an ordered set of $n$ Grassmann variables.
For any tensor $T$ of rank $n$ define a {\it generating function}
$T\in \calG(\theta)$ according to
\[
T(\theta)=\sum_{x\in \{0,1\}^n} T(x)\, \theta(x).
\]
Here
$\theta(x)=\theta_1^{x_1} \cdots \theta_n^{x_n}$ is the normally ordered monomial
corresponding to the subset of indexes  $x=\{ a\, : \, x_a=1\}$.
Let us introduce a linear differential operator $\Lambda$ acting on the tensor product
of two Grassmann algebras
$\calG(\theta)\otimes \calG(\theta)$ such that
\be
\label{Lambda_ad}
\Lambda = \sum_{a=1}^n  \theta_a \otimes \partial_a + \partial_a \otimes \theta_a.
\ee
\begin{lemma}
\label{lemma:matchgate2}
A tensor $T$ of rank $n$ is a matchgate iff
\be
\label{matchgate_identities2}
\Lambda \cdot T\otimes T=0.
\ee
\end{lemma}
\begin{proof}  For any strings $x,y\in \{0,1\}^n$ one has the following identity:
\[
(\theta_a \otimes \partial_a + \partial_a \otimes \theta_a)\cdot \theta(x)\otimes \theta(y)
=
\left\{
\ba{rcl}
0 &\mbox{if} & x_a=y_a,\\
 (-1)^{x_1+\ldots + x_{a-1} + y_1+\ldots + y_{a-1}}\, \theta(x\oplus e_a)\otimes \theta(y\oplus e_a)
&\mbox{if} & x_a\ne y_a.\\
\ea\right.
\]
Expanding both factors $T$ in Eq.~(\ref{matchgate_identities2}) in the monomials $\theta(x)$, $\theta(y)$,
using the above identity, and performing a change of variable $x\to x\oplus e_a$ and
$y\to y\oplus e_a$ for every $a$ one gets a linear combination of monomials
$\theta(x)\otimes \theta(y)$ with the coefficients given by the right hand side of
Eq.~(\ref{matchgate_identities}). Therefore Eq.~(\ref{matchgate_identities2}) is equivalent to Eq.~(\ref{matchgate_identities}).
\end{proof}
Lemma~\ref{lemma:matchgate2} provides an alternative definition of a matchgate tensor which is much
more useful than the original definition Eq.~(\ref{matchgate_identities}).
For example, it is shown below that the operator $\Lambda$ has a lot of symmetries which can be translated into
a group of transformations preserving the subset of matchagate tensors.
\begin{lemma}
\label{lemma:inv}
The operator $\Lambda$ is invariant under linear reversible changes of variables
$\theta_a=\sum_{b=1}^n U_{a,b}\, \tilde{\theta_b}$.
\end{lemma}
\begin{proof}
Indeed, let $\tilde{\partial}_a$ be the partial derivative over $\tilde{\theta}_a$.
Using Eq.~(\ref{derivative_change}) one gets
\[
\sum_{a=1}^n \theta_a \otimes \partial_a + {\partial}_a\otimes {\theta}_a =
\sum_{a,b,c=1}^n U_{a,b}\, (U^{-1})_{c,a} \, (\tilde{\theta}_b \otimes \tilde{\partial}_c +
 \tilde{\partial}_c \otimes \tilde{\theta}_b) =
\sum_{b}^n \, (\tilde{\theta}_b \otimes \tilde{\partial}_b + \tilde{\partial}_b \otimes \tilde{\theta}_b).
\]
\end{proof}
Lemmas~\ref{lemma:matchgate2},\ref{lemma:inv} imply that
linear reversible change of variables $T(\theta)\to T(\tilde{\theta})$,
where $\theta_a=\sum_{b=1}^n U_{a,b}\, \tilde{\theta_b}$ map matchgates to matchgates.
\begin{cor}
\label{cor:inv}
Let $T$ be a matchgate tensor of rank $n$.
Then a tensor $T'$ defined by any of the following transformations is also matchgate.\\
({\it Cyclic shift}):  $T'(x_1,x_2,\ldots,x_n)=T(x_2,\ldots,x_n,x_1)$,\\
({\it  Reflection}): $T'(x_1,x_2,\ldots,x_n)=T(x_n,\ldots,x_2,x_1)$,\\
({\it Phase shift}): $T'(x)=(-1)^{x\cdot z}\, T(x)$, where $z\in \{0,1\}^n$.\\
\end{cor}
\begin{proof}
Let $\epsilon=0$ if $T$ is an even tensor and $\epsilon=1$ if $T$ is an odd tensor,
see Corollary~\ref{cor:parity}. The transformations listed above
are generated  by the following linear changes of variables:
\bea
\mbox{\it Phase shift} &\mbox{:} & \theta_a \to (-1)^{z_a}\, \theta_a, \quad a=1,\ldots,n.\nn \\
\mbox{\it Cyclic shift} &\mbox{:} & \theta_a\to \theta_{a-1} \quad a=2,\ldots,n,
 \quad \mbox{and} \quad \theta_1\to (-1)^{\epsilon+1}\, \theta_n. \nn \\
 \mbox{\it Reflection} &\mbox{:} & \theta_a \to i\, \theta_{n-a}.\nn
\eea
Indeed, let $\theta(x)$ be the normally ordered monomial where $x=(x_1,x_2,\ldots,x_n)$.
Let  $x'=(x_2,\ldots,x_n,x_1)$ for the cyclic shift
and $x'=(x_n,\ldots,x_2,x_1)$ for the reflection. Then
the linear changes of variables stated above map $\theta(x)$ to
$(-1)^{z\cdot x}\, \theta(x)$ for the phase shift, to $\theta(x')$ for the cyclic shift, and
to $i^\epsilon\, \theta(x')$ for the reflection.
Therefore, in all three cases $T'$ is a matchgate tensor.
\end{proof}

\subsection{Matchgate tensors have Gaussian generating function}
\label{subs:matchgate=gaussian}
A memory size required to store a tensor of rank $n$ typically grows exponentially with $n$.
However the following theorem shows that for matchgate tensors the situation is much better.
\begin{theorem}
\label{thm:canonical}
A tensor $T$ of rank $n$ is a matchgate iff there exist an integer $0\le k\le n$, complex matrices
$A$, $B$ of size $n\times n$ and $k\times n$ respectively, and a complex number $C$ such that
$T$ has generating function
\be
\label{canonical}
T(\theta)=C\exp{\left( \frac12\, \theta^T\, A \, \theta \right)}\int D\mu\, \exp{\left( \mu^T \, B\, \theta \right)},
\ee
where $\mu=(\mu_1,\ldots,\mu_k)$ is a set of $k$ Grassmann variables. Furthermore, one can always choose
the matrices $A$ and $B$ such that $A^T=-A$ and $BA=0$.
\end{theorem}
Thus the triple $(A,B,C)$ provides a concise description of a matchgate tensor that
requires a memory size only $O(n^2)$.
In addition, it will be shown that contraction of matchgate tensors can be efficiently implemented using the
representation Eq.~(\ref{canonical}) and the Gaussian integral formulas of Section~\ref{subs:Gintegral}.
We shall refer to the generating function Eq.~(\ref{canonical}) as a {\it canonical generating function}
for a matchgate tensor $T$.
\begin{cor}
\label{cor:GI=matchgate}
For any matrices $A$ and $B$ the
 Gaussian integral $I(A,B)$ defined in Eq.~(\ref{Gaussian_Integrals}) is a matchgate.
 \end{cor}
 \begin{proof}
 Indeed, use  Eq.~(\ref{GI3}) and Theorem~\ref{thm:canonical}.
 \end{proof}
 In the rest of the section we shall prove Theorem~\ref{thm:canonical}.
\begin{proof}[Proof of Theorem~\ref{thm:canonical}.]
Let us first verify that the tensor defined in Eq.~(\ref{canonical}) is a matchgate,
i.e., $\Lambda\cdot T\otimes T=0$, see Lemma~\ref{lemma:matchgate2}.
Without loss of generality $A$ is an antisymmetric matrix and $C=1$.
Write $T$ as
\[
T=T_2\, T_1, \quad \mbox{where} \quad T_2=\exp{\left( \frac12\, \theta^T\, A \, \theta \right)},
\quad T_1=\int D\mu\, \exp{\left( \mu^T \, B\, \theta \right)}.
\]
Noting that  $T_2$ is an even function and
$\partial_a\, \theta(x) =\partial_a \cdot \theta(x) + \theta(x)\, \partial_a$ for any even string $x$
one concludes that
\be
\label{T1T2}
\Lambda\cdot T\otimes T=\left( \Lambda\cdot T_2\otimes T_2\right)\, T_1\otimes T_1+
T_2\otimes T_2\, \left( \Lambda\cdot T_1\otimes T_1\right).
\ee
Therefore it suffices to prove that $\Lambda\cdot T_2\otimes T_2=0$ and
$\Lambda\cdot T_1\otimes T_1=0$.
The first identity follows from $\partial_a\cdot T_2=\sum_{b=1}^n A_{a,b}\, \theta_b \, T_2$ and $A^T=-A$ which implies
\[
\Lambda\cdot T_2\otimes T_2=\sum_{a,b=1}^n A_{a,b}  \, (\theta_a\otimes\theta_b+ \theta_b\otimes
\theta_a)\,  T_2\otimes T_2 =0.
\]
To prove the second identity consider the singular value decomposition $B=L^T\tilde{B} R$, where $L\in SU(k)$ and $R\in SU(n)$ are unitary operators,  while $\tilde{B}$ is a $k\times n$ matrix with all non-zero elements located on the main diagonal,
$\tilde{B}=\mbox{diag}{(B_1,\ldots,B_k)}$. Introducing new variables $\tilde{\theta}=R\, \theta$
and $\tilde{\mu}=L\, \mu$ one gets
\[
T_1=\int D\tilde{\mu}\,  \exp{\left( \sum_{a=1}^k B_a\, \tilde{\mu}_a \, \tilde{\theta_a}\right)}=
B_1\cdots B_k\, \tilde{\theta}_1\cdots \tilde{\theta}_k.
\]
Here we have used identity $\int D\tilde{\mu}=\det{(L)}\, \int D\mu =\int D\mu$, see Eq.~(\ref{int_change}).
Since $\Lambda$ is invariant under linear reversible changes of variables,
see Lemma~\ref{lemma:inv}, and since
$\Lambda\cdot \theta(x)\otimes\theta(x)=0$ for any monomial $\theta(x)$
one gets $\Lambda\cdot T_1\otimes T_1=0$.
We proved that $\Lambda\cdot T\otimes T=0$, that is, $T$ is a matchgate tensor.

Let us now show that any matchgate tensor $T$ of rank $n$  can be written as in Eq.~(\ref{canonical}).
Define a linear subspace $\calZ\subseteq \CC^n$ such that
\[
\calZ=\{ \xi \in \CC^n \, : \, \sum_{a=1}^n \xi_a \theta_a T=0\}.
\]
Let $\dim{(\calZ)}=k$. Make a change of variables $\eta= U\, \theta$
where $U$ is any invertible matrix such that the last $k$ rows of $U$ span $\calZ$.
Then $\eta_a\, T=0$ for all $a=n-k+1,\ldots,n$. It follows that $T$ can be represented
as
\be
\label{T=}
T=\eta_{n-k+1} \cdots \eta_{n} \, S
\ee
for some function $S=S(\eta)$ that depends only on variables $\eta_1,\ldots,\eta_{n-k}$.
Equivalently,
\[
S=\partial_{n}\cdots \partial_{n-k+1} \cdot T,
\]
where the partial derivatives are taken with respect to the variables $\eta$.
Since $\Lambda$ is invariant under reversible linear changes of variables, see Lemma~\ref{lemma:inv},
and since $\Lambda\, \partial_a\otimes \partial_a = \partial_a\otimes \partial_a \, \Lambda$, we get
\be
\label{LambdaSS}
\Lambda \cdot S\otimes S=\sum_{a=1}^{n-k} \eta_a S \otimes \partial_a\cdot  S +
\partial_a \cdot S \otimes \eta_a S=0.
\ee
By definition of the subspace $\calZ$ the functions $\eta_1 S,\ldots,\eta_{n-k} S$ are linearly
independent. Therefore there exist linear functionals $F_a \, : \, \calG(\eta) \to \CC$, $a=1,\ldots,n-k$, such that
$F_a(\eta_b S)=\delta_{a,b}$. Applying $F_a$ to the first factor in  Eq.~(\ref{LambdaSS}) we
get
\be
\label{difur1}
\partial_a \cdot S =\sum_{b=1}^{n-k} M_{a,b}\, \eta_b\, S, \quad \mbox{where} \quad M_{a,b}=-F_a(\partial_b \cdot S)\in \CC,
\ee
for all   $a=1,\ldots,n-k$.
Let $k_{min}$  the lowest degree of monomials in $S$. Let us show that $k_{min}=0$,
that is, $S(\eta)$ contains $I$ with a non-zero coefficient. Indeed, let $S_{min}$ be a
function obtained from $S$ by retaining only
monomials of degree $k_{min}$. Since any monomial in the r.h.s. of Eq.~(\ref{difur1}) has degree
at least $k_{min}+1$, we conclude that $\partial_a \cdot S_{min}=0$ for all $a$. It means that
$S_{min}=C\, I$ for some complex number $C\ne 0$ and thus $k_{min}=0$.

Applying the partial derivative $\partial_b$ to Eq.~(\ref{difur1})
we get $M_{a,b} =C^{-1} (\partial_b\, \partial_a \cdot \left. S)\right|_{\eta=0}$, where the substitution $\eta=0$
means that the term proportional to the identity is taken. Since the partial derivatives over different variables
anticommute, $M$ is an antisymmetric   matrix.

Using Gaussian elimination
any antisymmetric matrix $M$ can be brought into a block-diagonal form with $2\times 2$ blocks
on the diagonal by a transformation $M\to M'=W^T\, X \, W$, where $W$ is an invertible
matrix (in fact, one can always choose unitary $W$, see~\cite{Zumino62}).
Since our change of variables $\eta=U\theta$ allows arbitrary  transformations
in the subspace of $\eta_1,\ldots,\eta_{n-k}$ we can assume that $M$ is already bock-diagonal,
\[
M=\bigoplus_{a=1}^m \left( \ba{cc} 0 & \lambda_a \\ -\lambda_a & 0 \\ \ea \right), \quad \lambda_1,\ldots,\lambda_m\in \CC,
\]
where only non-zero blocks are represented, so that $2m\le n-k$.

Applying  Eq.~(\ref{difur1}) for $a=1,2$ we get
\be
\label{difur2}
\partial_1 \cdot S = \lambda_1 \eta_2 S, \quad \partial_2\cdot S = -\lambda_1 \eta_1 S.
\ee
Note that  $S$ can be written as
\be
\label{expand}
S=\sum_x (\alpha_x \eta_1 + \beta_x \eta_2)\eta(x) + \sum_y (\gamma_y I + \delta_y \eta_1\eta_2) \eta(y),
\ee
where the sums over $x$ and $y$ run over all odd and even monomials in $\eta_3,\ldots,\eta_{n-k}$ respectively.
Substituting Eq.~(\ref{expand}) into Eq.~(\ref{difur2}) one gets $\alpha_x=\beta_x=0$ and
$\delta_x=\lambda_1 \gamma_x$, that is
\[
S=(I+\lambda_1 \eta_1\eta_2) S',
\]
where $S'$ depends only on variables $\eta_3,\ldots,\eta_{n-k}$. Repeating this
argument inductively, we arrive to the representation
\[
S=C\, \prod_{a=1}^m (I+\lambda_a \eta_{2a-1}\eta_{2a}) = C\, \exp{\left( \frac12 \eta^T\, M \, \eta\right)}.
\]
Here we extended the matrix $M$ such that its last $k$ columns and rows are zero.
Combining it with Eq.~(\ref{T=}) one gets
\[
T=C\, \eta_{n-k+1} \cdots \eta_n \, \exp{\left( \frac12 \eta^T\, M \, \eta\right)}=C\,
\exp{\left( \frac12 \eta^T\, M \, \eta\right)}\,
\int D\mu\, \exp{\left( \mu^T \, \tilde{B}\, \eta \right)},
\]
where $\mu$ is a vector of $k$ Grassmann variables  and $\tilde{B}$ is a $k\times n$ matrix with $0$,$1$ entries
such that
\[
\mu^T \, \tilde{B}\, \eta = \sum_{a=1}^k \mu_a\, \eta_{n-k+a}.
\]
Recalling that $\eta=U\, \theta$, we conclude that $T$ has a representation Eq.~(\ref{canonical})
with $A=U^T\, M\, U$ and $B=\tilde{B}\, U$.
As a byproduct we also proved that the matrices $A$, $B$ in Eq.~(\ref{canonical}) can
always be chosen such that $BA=0$ since $BA=\tilde{B}\, M\, U$ and all non-zero entries
of $\tilde{B}$ are in the last $k$ rows.
\end{proof}

\subsection{Graph theoretic definition of matchgate tensors}
\label{subs:matchgate=matchsum}

Let $G=(V,E,W)$ be an arbitrary weighted graph with a set of vertices $V$,
set of edges $E$ and a weight function  $W$ that assigns a complex weight $W(e)$
to every edge $e\in E$.
\begin{dfn}
Let $G=(V,E)$ be a graph and $S\subseteq V$ be a subset of vertices.
A subset of edges $M\subseteq E$ is called an $S$-imperfect matching iff every vertex from $S$
has no incident edges from $M$ while every vertex from $V\backslash S$ has
exactly one incident edge from $M$. A set of all $S$-imperfect matchings in a graph $G$ will be denoted
 $\calM(G,S)$.
\end{dfn}
Note that a perfect matching corresponds to an $\emptyset$-imperfect matching.
Occasionally we shall denote a set of perfect matching by $\calM(G)\equiv \calM(G,\emptyset)$.
For any subset of vertices $S\subseteq V$ define a {\it matching sum}
\be
\label{Zcycles1}
\perfectm{G}{S}=\sum_{M\in \calM(G,S)} \; \prod_{e\in M} W(e).
\ee
(A matching sum can be identified with a planar matchgate of~\cite{Valiant07}.)
In this section we outline an isomorphism between matchgate tensors
and matching sums of planar graphs discovered earlier in~\cite{Cai2006c}.
For the sake of completeness we provide a proof of this result below.
Although the main idea of the proof is the same as in~\cite{Cai2006c} some technical
details are different. In particular, we use much simpler crossing gadget.

Specifically, we shall consider planar weighted graphs $G=(V,E,W)$ embedded into a disk
such that some subset of $n$ {\it external} vertices $V_{ext}\subseteq V$ belongs to the boundary of disk
while all other {\it internal} vertices $V\backslash V_{ext}$  belong to the interior of $D$.
Let $V_{ext}=\{u_1,\ldots,u_n\}$ be an ordered list of external vertices  corresponding to circumnavigating
anticlockwise the boundary of the disk.
Then any binary string $x\in \{0,1\}^n$ can be identified with a subset
$x\subseteq V_{ext}$ that includes all external vertices $u_j$ such that $x_j=1$.
Now we are ready to state the main result of this section.
\begin{theorem}
\label{thm:cycles}
For any matchgate tensor $T$ of rank $n$ there exists a planar weighted graph $G=(V,E,W)$
with $O(n^2)$ vertices,  $O(n^2)$ edges  and a subset of $n$ vertices $V_{ext}\subseteq V$
such that
\be
\label{T=Z}
T(x)=\perfectm{G}{x} \quad \mbox{for all} \quad x\subseteq V_{ext}.
\ee
Furthermore, suppose $T$ is specified by its generating function,
$T=C\exp{\left( \frac12\, \theta^T\, A \, \theta \right)}\int D\mu\, \exp{\left( \mu^T \, B\, \theta \right)}$.
Then the graph $G$ can be constructed in time $O(n^2)$
and the weights $W(e)$ are linear functionals  of $A$, $B$, and $C$.
\end{theorem}
The key step in proving the theorem is to show that  Pfaffian of any $n\times n$  antisymmetric matrix
can be expressed as a matching sum  on some planar graph with
$O(n^2)$ vertices. This step can be regarded as a reversal of the FKT method that allows one to
represent the matching sum of a planar graph as Pfaffian of the Tutte matrix.
\begin{lemma}
\label{lemma:Pf=cycles}
For any complex antisymmetric matrix $A$ of size $n\times n$ there exists a planar weighted graph $G=(V,E,W)$
 with $O(n^2)$ vertices, $O(n^2)$ edges such that
 the weights $W(e)$  are linear functionals  of $A$ and
 \be
 \label{Pf=cycles}
 \pf{(A)}=\perfectm{G}{\emptyset}.
 \ee
 The graph $G$ can be constructed in time $O(n^2)$.
 \end{lemma}
{\it Remark:} It should be emphasized that we regard both sides of Eq.~(\ref{Pf=cycles}) as
polynomial functions of matrix elements of $A$, and the lemma states that the two polynomials coincide.
However, even if one treats both sides of Eq.~(\ref{Pf=cycles}) just as complex numbers, the statement
of the lemma is still non-trivial, since one can not compute $\pf{(A)}$ in time $O(n^2)$ and thus one
has to construct the graph $G$ without access to the value of $\pf{(A)}$.
\begin{proof}
Let us assume that $n$ is even (otherwise the statement is trivial).
Let $D$ be a disk with $n$ marked points $v_1,\ldots,v_n$ on the boundary such that  their order corresponds
to anticlockwise circumnavigating the boundary of $D$. Let $C_n$ be the complete graph with vertices
$v_1,\ldots,v_n$ embedded into $D$. We assume that the embedding is chosen such that all edges of $C_n$
lie inside the disk and there are only double edge crossing points, see Fig.~1.
\begin{figure}
\label{fig:complete}
\centerline{
\mbox{
 \includegraphics[height=3cm]{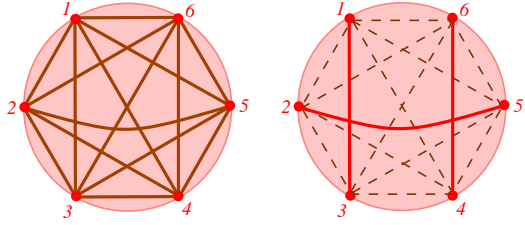}
 }}
\caption{Left: a complete graph $C_6$ embedded into a disk. Right: a perfect matching on $C_6$ with two
self-intersections.}
\end{figure}
Let $\calM(C_n)$ be a set of perfect matchings on $C_n$. For any perfect matching $M\in \calM(C_n)$
let $N_c(M)$ be the number of self-intersections in $M$, i.e., the number of edge crossing points
in the planar embedding of $C_n$ in which
both crossing edges  are occupied by $M$. For example, given a planar embedding of $C_6$ shown on
Fig.~1, a perfect matching $M=(1,3),(2,5),(4,6)$ has two self-intersections.
We claim that
\be
\label{pf-crossing}
\pf{(A)}=\sum_{M\in \calM(C_n)} (-1)^{N_c(M)}\, \prod_{(u,v)\in M,\; u<v} A_{u,v}.
\ee
Indeed, by definition of  Pfaffian
\be
\label{pf-standard}
\pf{(A)}=\sum_{\sigma} \mathrm{sgn}(\sigma) \, A_{\sigma(1),\sigma(2)} \cdots A_{\sigma(n-1),\sigma(n)},
\ee
where the sum is over all permutations of $n$ elements $\sigma$ such that
$\sigma(2j-1)<\sigma(2j)$ for all $j$ and $\sigma(1)<\sigma(3)<\ldots <\sigma(n-1)$.
Clearly, there exists a one-to-one correspondence between such permutations and perfect matchings in $C_n$.
If $M$ is the perfect matching corresponding to the identity permutation, $M=(1,2),\ldots,(n-1,n)$,
one has $N_c(M)=0$ and the signs in Eqs.~(\ref{pf-crossing},\ref{pf-standard}) coincide. Furthermore, changing $M$ by
any transposition $j\leftrightarrow j+1$ either does not change $M$ or
changes the parity of $N_c(M)$, so the signs in Eqs.~(\ref{pf-crossing},\ref{pf-standard})
coincide for all perfect matchings.

In order to represent the sum over perfect matchings in Eq.~(\ref{pf-crossing}) as a sum over perfect matchings
in a planar graph
we shall replace each edge crossing point of $C_n$ by a {\it crossing gadget}, see Fig.~2.
A {\it crossing gadget} is a planar simulator
for an edge crossing point. It allows one to establish a correspondence between subsets of edges
in the non-planar graph  and subsets of edges in a planar graph. In addition, a crossing gadget
will take care of the extra sign\footnote{One can gain some intuition about the extra sign factor in Eq.~(\ref{pf-crossing})
if one thinks about the set of edges occupied by a perfect matching $y$ as a family of
"world lines" of fermionic particles.  The contribution from $y$ to $\pf{(A)}$ can be thought of as a quantum amplitude
assigned to this family of world lines. Whenever two particles are exchanged the amplitude
acquires an extra factor $-1$.} factor in Eq.~(\ref{pf-crossing}).

\noindent
{\it Crossing gadget.} Consider a weighted graph $G_{cross}$ shown on Fig.~2.
It has $6$ vertices and $7$ edges. The edge $(5,6)$ carries weight $-1$ and all other edges
carry weight $+1$.
We fix the embedding of $G_{cross}$ into a disk such that $G_{cross}$ has four external vertices $\{1,2,3,4\}$
on the boundary of the disk.
One can easily check that the matching sum of $G_{cross}$ satisfies the following identities:
\bea
\perfectm{G_{cross}}{\emptyset}&=&1,\nn \\
\perfectm{G_{cross}}{\{1,3\}}=\perfectm{G_{cross}}{\{2,4\}}&=&1, \nn \\
\perfectm{G_{cross}}{\{1,2,3,4\}}&=&-1, \nn \\
\perfectm{G_{cross}}{\{1,2\}}=\perfectm{G_{cross}}{\{3,4\}}&=&0, \nn \\
\perfectm{G_{cross}}{\{1,4\}}=\perfectm{G_{cross}}{\{2,3\}}&=&0. \nn
\eea
These identities are illustrated in Fig.~3.
In addition, $\perfectm{G_{cross}}{S}=0$ whenever $|S|$ is odd.
Thus the four boundary conditions for which the matching sum is non-zero
represents the four possible configurations (empty/occupied) of a pair of  crossing edges
if they were attached to the vertices $\{1,2,3,4\}$.
For every edge crossing point of $C_n$ one has to cut out
a small disk centered at the crossing point and replace the interior of the disk by the gadget $G_{cross}$ such that the
four vertices $\{1,2,3,4\}$ are attached to the four external edges, see Fig.~2. Let $\tilde{C}_n$ be the resulting graph.
By construction, $\tilde{C}_n$ is planar.  It remains to assign weights to edges of $\tilde{C}_n$
such that
\be
\label{correct_weights}
\perfectm{\tilde{C}_n}{\emptyset}=\pf{(A)}.
\ee

\begin{figure}
\label{fig:crossing}
\centerline{
\mbox{
 \includegraphics[height=3cm]{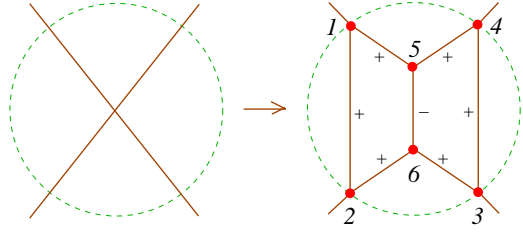}
 }}
\caption{Each edge crossing point in the
planar embedding of the complete graph $C_n$ is replaced by the crossing gadget $G_{cross}$.
Edges labeled by $\pm $ carry a weight $\pm 1$.}
\end{figure}

\begin{figure}
\label{fig:crossing1}
\centerline{
\mbox{
 \includegraphics[height=4cm]{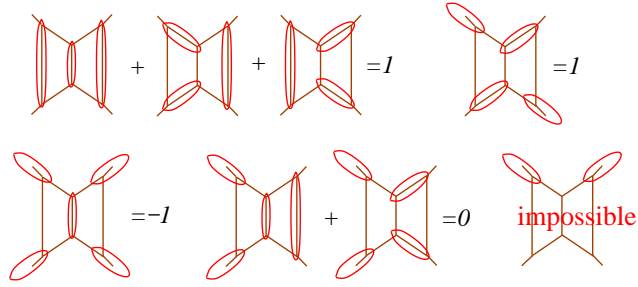}
 }}
\caption{Matching sums of the graph $G_{cross}$ corresponding to various boundary conditions.}
\end{figure}

Any edge of $\tilde{C}_n$ falls into one of the four categories:
(i) edge of $C_n$;
(ii) a section of some edge of  $C_n$ between two crossing gadgets;
(iii) a section of some edge of $C_n$ between a vertex of $C_n$ and some crossing gadget;
(iv) an edge that belongs to some crossing gadget. Note that the edges of type (iv) have been already assigned
a weight, whereas any edge of type (i),(ii), and (iii) has a unique ancestor edge $e=(u,v)$ in $C_n$.
 Let us agree that for every edge $e=(u,v)$, $u<v$ of $C_n$  we choose one of its
descendants $\tilde{e}$ in $\tilde{C}_n$  and assign $\tilde{e}$ the weight $A_{u,v}$, while all other descendants of $e$ are assigned the weight $1$.
Since all descendants of $e$ appear or do not appear in any perfect matching $M\in \calM(\tilde{C}_n)$
simultaneously,
we arrive to Eq.~(\ref{correct_weights}), that is, $\tilde{C}_n$ is the desired graph $G$.
It remains to count the number of vertices in $\tilde{C}_n$. There are $O(n^2)$ crossing gadgets each having $O(1)$
vertices. Thus $\tilde{C}_n$ has $O(n^2)$
vertices. Since $\tilde{C}_n$ is a planar graph it has $O(n^2)$ edges, see~\cite{Diestel}.
\end{proof}

Let $\tilde{C}_n$ be a planar graph constructed above. Consider a matching sum
$\perfectm{\tilde{C}_n}{S}$ for some subset $S\subseteq \{1,2,\ldots,n\}$ of vertices
lying on the boundary of the disk. By repeating the arguments used in the proof of Lemma~\ref{lemma:Pf=cycles}
one concludes that
\be
\label{Pf=cycles1}
\perfectm{\tilde{C}_n}{S} = \pf{(A[S])} \quad \mbox{for all} \quad S\subseteq \{1,2,\ldots,n\},
\ee
where  $A[S]$ is a matrix obtained from $A$ by removing all rows and columns $a\in S$.
Theorem~\ref{thm:cycles} follows from Eq.~(\ref{Pf=cycles1}) and the following simple observation.
\begin{lemma}
\label{lemma:T=pfaffian}
Let $T$ be a matchgate tensor of rank $n$  with a parity $\epsilon(T)$ specified by its generating function
\be
\label{T=GIntegral}
T=C\exp{\left( \frac12\, \theta^T\, F\, \theta \right)}\int D\mu\, \exp{\left( \mu^T \, G\, \theta \right)}.
\ee
Then
\be
\label{T=pfaffian}
T(x)=C \epsilon(T)\, \pf{\left( A(x\, 1^{k-n})\right)} \quad \mbox{for all $x\in \{0,1\}^n$},
\quad \mbox{where} \quad
A=\left[ \ba{cc}
F & -G^T \\
G & 0 \\
\ea
\right].
\ee
The matrix $A$ has size $k\times k$ with $n\le k\le 2n$.
\end{lemma}
\noindent
{\it Remark:} As usual,
$A(y)$ denotes a matrix obtained from $A$ by removing all columns and rows $a$ such that $y_a=0$.
We assume that $\epsilon(T)=1$ ($\epsilon(T)=-1$)
for even (odd) tensors.
\begin{proof}
Theorem~\ref{thm:canonical} asserts that $T$ always has a generating
function Eq.~(\ref{T=GIntegral}) where $G$ has size $m\times n$ for some $m\le n$.
Thus $k=n+m\le 2n$.
Introducing  a set of $k$ Grassmann variables $\eta=(\theta_1,\ldots,\theta_n,\mu_1,\ldots,\mu_m)$ one can rewrite $T$ as
\[
T(\theta)=C\, \int D\mu \exp{\left( \frac12\, \eta^T\, A\, \eta \right)}.
\]
Expanding the exponent  one gets
\[
\exp{\left( \frac12\, \eta^T\, A\, \eta \right)}=\sum_{z\in \{0,1\}^k}
\pf{(A(z)})\,  \eta(z).
\]
Note that
\[
\int \, D\mu \, \eta(z)=\left\{ \ba{rcl}
(-1)^{m(z_1+\cdots+z_n)} &\mbox{if} & z_{n+1}=\ldots=z_k=1,\\
0 &&\mbox{otherwise}.\\ \ea \right.
\]
Taking into account that $m$ is even (odd) for even (odd) tensors and so is
$z_1+\cdots +z_n$ we conclude that
\be
\label{T=sum}
T(\theta)=C \epsilon(T)\, \sum_{x\in \{0,1\}^n}
\pf{(A(x1^{k-n})})\, \theta(x),
\ee
that is $T(x)=C \epsilon(T)\, \pf{(A(x1^{k-n}))}$.
\end{proof}
\begin{proof}[Proof of Theorem~\ref{thm:cycles}]
Let $A$ be the $k\times k$ matrix constructed in Lemma~\ref{lemma:T=pfaffian}
and $\tilde{C}_k$ be the weighted planar graph constructed in Lemma~\ref{lemma:Pf=cycles}
such that Eq.~(\ref{Pf=cycles1}) holds for all $S\subseteq \{1,2,\ldots,k\}$.
Therefore,
\be
\label{aux314}
T(x)=C\epsilon(T)\, \perfectm{\tilde{C}_k}{\bar{x}0^{k-n}} \quad \mbox{for all} \quad  x\in \{0,1\}^n
\ee
where $\bar{x}$ is obtained from $x$ by flipping every bit.
In order to transform Eq.~(\ref{aux314}) into Eq.~(\ref{T=Z}) one can
incorporate the factor $C\epsilon(T)$ into the matching sum by introducing an extra
edge with a weight $C\epsilon(T)$ and adding one extra edge with weight $1$ to
every vertex $1,2,\ldots,n$ of the graph $\tilde{C}_k$ in order to flip bits of $x$.
\end{proof}

Although it is not necessary, let us mention that the
reverse of Theorem~\ref{thm:cycles} is also true, namely, a tensor $T$
defined by Eq.~(\ref{T=Z}) is always a matchgate. The easiest way to prove it is to
represent the matching sum $\perfectm{G}{x}$ in Eq.~(\ref{T=Z}) as a contraction
of an open matchgate tensor network, see Section~\ref{subs:1shot},
in which every tensor has a linear generating function (thus simulating the perfect
matching condition).  Then one can use Corollary~\ref{cor:matchgate} to prove that
$T$ is a matchgate.

\section{Contraction of matchgate tensor networks}

\subsection{Edge contractions}
Consider a tensor network $\calT$ defined on a graph $G=(V,E)$ embedded to a surface $\Sigma$.
Suppose one can find a region $D\subset \Sigma$  with  topology of a disk
such that $D$ contains exactly two vertices $u,v\in V$
and several edges connecting $u$ and $v$ as shown on Fig.~4.
We shall define a new tensor network $\calT'$
such that: (i) $\calT'$ coincides with $\calT$ outside $D$; (ii)  $\calT'$ contains only one vertex inside $D$;
(iii) contraction values of $\calT$ and $\calT'$ are the same.
The operation of replacing $\calT$ by $\calT'$ will be referred to as an {\it edge contraction}.
The new vertex obtained by contracting all edges connecting $u$ and $v$ inside $D$ will be denoted $u\star v$.

\begin{figure}
\label{fig:edge_contraction}
\centerline{
\mbox{
 \includegraphics[height=3cm]{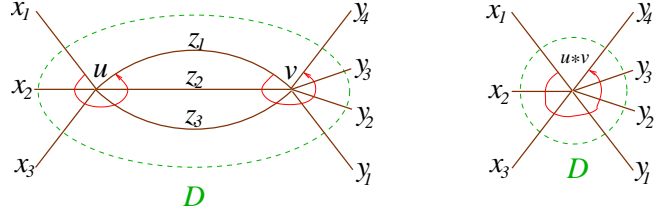}
 }}
\caption{The ordering of edges before and after contraction of $u$ and $v$.}
\end{figure}

Suppose there are $b$ edges connecting $u$ and $v$ that lie inside the disk.
Applying, if necessary, a cyclic shift of components to the tensors $T_u$ and/or $T_v$ we can assume that these edges
correspond to the last $b$ components of the tensor $T_u$ and the first $b$ components of $T_v$, see
Fig.~4.
Note that if the tensors under consideration are matchgates, the
tensors obtained after the cyclic shift are also matchgates, see Corollary~\ref{cor:inv}.
In the new network $\calT'$  a pair of vertices $u,v$ is replaced by a single vertex $u\star v$ with degree
$d(u\star v)=d(u)+d(v)-2b$.
We define a new tensor $T_{u\star v}$  as
\be
\label{edge_contraction}
T_{u\star v}(x,y)=\sum_{z_1,\ldots,z_b=0,1} T_u(x,z_b,z_{b-1},\ldots,z_1)\, T_v(z_1,\ldots,z_{b-1},z_b,y),
\ee
where $x$ and $y$ can be arbitrary bit strings of length $d(u)-b$ and $d(v)-b$ respectively.
By definition of the contraction value, $c(\calT)=c(\calT')$.

We shall also define a {\it self-loop contraction} as a special case of edge contraction. Namely,
suppose one can find a region $D\subset \Sigma$ with topology of a disk
such that $D$ contains exactly one vertex $u\in V$
and several self-loops as shown on Fig.~5.  We shall define a new tensor network $\calT'$
such that: (i) $\calT'$ coincides with $\calT$ outside $D$; (ii)  $\calT'$ contains one vertex without self-loops
inside $D$; (iii) contraction values of $\calT$ and $\calT'$ are the same.
The operation of replacing $\calT$ by $\calT'$ will be referred to as a {\it self-loop contraction}.
To define this operation,
choose  the most inner self-loop $\gamma \in E(u)$ introduce a dummy vertex $v$ near the median of $\gamma$
and assign a tensor $T_v(x_1,x_2)=\delta_{x_1,x_2}$ to this vertex. Clearly it does not change a contraction value of a network.
Secondly, apply the edge contraction described above to the two edges connecting $u$ and $v$. This reduces the number
of self-loops by one. Repeat these two steps until all self-loops inside $D$ are contracted.

\begin{figure}
\label{fig:loop_contraction}
\centerline{
\mbox{
 \includegraphics[height=2cm]{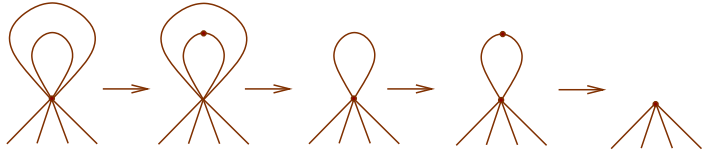}
 }}
\caption{Contraction of self-loops can be reduced to edge contraction by adding dummy vertices.}
\end{figure}

It should be mentioned that self-loops $\gamma\in E(u)$ can be identified with elements of the
fundamental group $[\gamma]\in \pi_1(\Sigma,u)$ of the surface $\Sigma$ with a base point $u$. We do not allow to contract
self-loops representing non-trivial homotopy classes (because it cannot  be done efficiently for
matchgate tensor networks).

\subsection{Edge contraction as a convolution of  generating functions}
\label{subs:econtraction}
Let $\calT=\{T_u\}_{u\in V}$ be a tensor network considered in the previous section.
In order to describe each tensor $T_u$ by a generating function $T_u(\theta)$
we shall introduce Grassmann variables $\theta_{u,1},\ldots,\theta_{u,d(u)}$
associated with the edges $e^u_1,\ldots,e^u_{d(u)}\in E(u)$ incident to $u$ such that
\be
\label{T_u(theta)}
T_u(\theta)=\sum_{x\in \{0,1\}^n} T(x)\, (\theta_{u,1})^{x_1} (\theta_{u,2})^{x_2}  \cdots (\theta_{u,n})^{x_n},
\quad n\equiv d(u).
\ee
Similarly one can describe the contracted tensor $T_{u\star v}$ in Eq.~(\ref{edge_contraction})
by  a generating function
\be
\label{Tu*v_def}
T_{u\star v}(\theta)=\sum_{x\in \{0,1\}^p} \sum_{y\in \{0,1\}^q}
T_{u\star v}(x,y) \, (\theta_{u,1})^{x_1} \cdots (\theta_{u,p})^{x_p} (\theta_{v,b+1})^{y_1} \cdots (\theta_{v,b+q})^{y_q},
\ee
where $p\equiv d(u)-b$ and $q\equiv d(v)-b$.
The goal of this section is to represent the function $T_{u\star v}(\theta)$
as an integral of $T_{u}(\theta) T_{v}(\theta)$ in which all variables
associated with the edges to be contracted are integrated out.

 Let $E(u,v)$ be a set of edges connecting $u$ and $v$.
For any edge $e\in E(u,v)$ such that $e$ is labeled as $e_j^u\in E(u)$ and as $e_k^v\in E(v)$
denote
\[
\theta(e)=\theta_{u,j}\, \theta_{v,k}, \quad  \int\,  D\theta(e)=\int\,  d\theta_{v,k}\,\int \, d\theta_{u,j},
\quad \mbox{and} \quad \int_{e\in E(u,v)} D\theta(e) =\prod_{e\in E(u,v)}  \int\,  D\theta(e).
\]
Note that these definitions make sense only $(u,v)$ is regarded as an ordered pair of vertices.
Also note that the integrals $\int\, D\theta(e)$ over different edges commute, see Eq.~(\ref{int_com}),
and thus one can take the integrals in an arbitrary order.
\begin{lemma}
\label{lemma:convolution}
Suppose the edges connecting $u$ and $v$ are ordered as shown on Fig.~4, i.e.,
these are the last $b$ edges incident to $u$ and the first $b$ edges incident to $v$. Then
\be
\label{TuTv}
T_{u\star v}=\int_{e\in E(u,v)} D\theta(e)\; T_u\, T_v\, \exp{\left(
\sum_{e\in E(u,v)} \theta(e)\right)}.
\ee
\end{lemma}
\begin{proof}
By linearity it is enough to prove Eq.~(\ref{TuTv}) for the case when $T_u$ and $T_v$ are monomials
in the Grassmann variables, i.e.,
\[
T_u= (\theta_{u,1})^{x_1} \cdots (\theta_{u,p})^{x_p} (\theta_{u,p+1})^{z_1'} \cdots (\theta_{u,p+b})^{z_b'},
\quad
T_v= (\theta_{v,1})^{z_1} \cdots (\theta_{v,b})^{z_b} (\theta_{v,b+1})^{y_1} \cdots (\theta_{v,q+b})^{y_q},
\]
where $p\equiv d(u)-b$ and $q\equiv d(v)-b$. By expanding the exponent
one gets a sum of all possible monomials in which the two variables associated with any edge $e\in E(u,v)$
are either both present or both absent.
Therefore the integral in Eq.~(\ref{TuTv}) is zero unless $z_j=z_{b+1-j}'$ for all $j=1,\ldots,b$. Suppose this is the case.
Then one gets after some rearrangement of variables
\[
T_u\, T_v= (\theta_{u,1})^{x_1} \cdots (\theta_{u,p})^{x_p} \left( \prod_{e\in S(z)} \theta(e) \right)
(\theta_{v,b+1})^{y_1} \cdots (\theta_{v,d(v)})^{y_q},
\]
where $S(z)\subseteq E(u,v)$ denotes a set of edges $e$ such that $e$ is labeled
as $e_k^v\in E(v)$ and $z_k=1$.
Substituting it into the integral Eq.~(\ref{TuTv}), taking into account that $\theta(e)$
is a central element and that $\int D\theta(e) \, \theta(e)=1$ one gets
\[
T_{u\star v}=(\theta_{u,1})^{x_1} \cdots (\theta_{u,p})^{x_p} \,
(\theta_{v,b+1})^{y_1} \cdots (\theta_{v,b+q})^{y_q}
\]
which coincides with the desired expression Eq.~(\ref{Tu*v_def}).
 \end{proof}

\begin{cor}
\label{cor:matchgate}
Suppose $T_u$ and $T_v$ are matchgates. Then the contracted tensor $T_{u\star v}$ is also a matchgate.
\end{cor}
\begin{proof}
Since cyclic shifts of indexes map matchgates to matchgates, see Corollary~\ref{cor:inv}
in Section~\ref{subs:generating},
we can assume that the edges of $T_u$ and $T_v$ are already ordered as required in
 Lemma~\ref{lemma:convolution}. Represent $T_u$, $T_v$ by their canonical
 generating functions, see Theorem~\ref{thm:canonical}.
Using  Eq.~(\ref{TuTv})  one concludes that $T_{u\star v}(\theta)$ is a Gaussian
integral $I(A,B)$ for some matrices $A$ and $B$, see Eq.~(\ref{Gaussian_Integrals}).
Therefore, $T_{u\star v}$ is a matchgate, see Corollary~\ref{cor:GI=matchgate}
in Section~\ref{subs:matchgate=gaussian}.
\end{proof}

{\it Remark:}  Given the canonical generating functions for $T_u$ and $T_v$,
the canonical generating function for the contracted tensor $T_{u\star v}$
can be obtained straightforwardly using Eq.~(\ref{TuTv}) and computing the
resulting Gaussian integral $I(A,B)$ using Eq.~(\ref{GI3}).  The details can be
found in Appendix~A.

\subsection{Contraction of a planar subgraph in one shot}
\label{subs:1shot}
Suppose a planar connected graph $G=(V,E)$ is a part of a larger non-planar tensor network
such that $G$ is connected to the rest of the network by a subset of {\it external edges} $E_{ext}\subseteq E$.
The remaining {\it internal edges} $E_{int}=E\backslash E_{ext}$ are the edges that can be
contracted "locally" without touching the rest of the network.
By abuse of definitions, we shall assume that the external edges have only one endpoint (the other endpoint
belongs to the rest of the network) which belongs to the outer face of $G$, see~Fig.~6.
For convenience let us also assume that the graph $G$ is embedded
into a disk such that the external edges stick out from the disk as shown on Fig.~6.
A network that consists of such a graph $G=(V,E)$ and a collection of
tensors $\{T_u\}_{u\in V}$ will be referred to as an
{\it open tensor network}.  Throughout this section we shall consider only open tensor networks
in which every tensor is a matchgate.
Contraction of an open tensor network  amounts to finding a tensor
$T_V$ of rank $|E_{ext}|$ obtained by contracting all internal edges of $G$.
It follows from Corollary~\ref{cor:matchgate}, Section~\ref{subs:econtraction} that $T_V$
is a matchgate.
The goal of the present section is to represent the generating function for the contracted tensor
$T_V$ as a convolution integral similar to Eq.~(\ref{TuTv}) where the integration is taken over all
internal edges.

\begin{figure}
\label{fig:planar_open}
\centerline{
\mbox{
 \includegraphics[height=4cm]{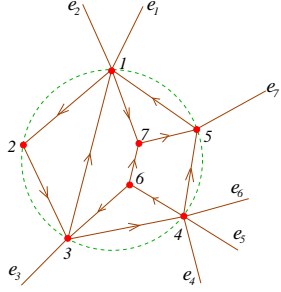}
 }}
\caption{An open tensor network with $7$ external edges equipped with a Kasteleyn orientation. }
\end{figure}

An alternative strategy for computing $T_V$ is to apply the edge contraction described in the previous section
sequentially until all internal edges of $G$ are contracted. Although it yields a polynomial-time
 algorithm this strategy is not very robust.
  An obvious drawback
is that every edge contraction involves computing the Gaussian integral Eq.~(\ref{GI3}) which requires
a matrix inversion. Contracting sequentially $O(n)$ edges would require $O(n)$ nested matrix
inversions  which may be difficult or impossible to do if the matrix elements are specified with a
finite precision.
In order to reduce the number of nested matrix inversions one could
organize the edge contractions into a sequence of rounds such that each round involves contractions
of pairwise disjoint edges. The contractions involved in every round can be performed in parallel.
The number of the rounds can be made $O(\log{n})$ using the techniques developed
by F\"urer and Raghavachari~\cite{Furer}.
We shall not pursue this strategy though because the approach described below allows one to compute $T_V$
using only  one matrix inversion.

The main result of this section is the following theorem.
\begin{theorem}
\label{thm:partial_contraction}
Consider an open matchgate tensor network
on a planar graph $G=(V,E)$ with $n$ vertices and $m$ external edges.
Assume that the  tensors $T_1,\ldots,T_n$ are specified by their canonical
generating function,
\[
T_j(\theta)=C_j\exp{\left( \frac12\, \theta^T\, A_j \, \theta \right)}\int D\mu\, \exp{\left( \mu^T \, B_j\, \theta \right)}.
\]
Then the tensor $T_V$ obtained by contracting all internal edges of $G$ can be represented as a Gaussian integral
\be
\label{integral_main}
T_V(\eta)= \prod_{j=1}^n C_j \epsilon(T_j)\, \int D\theta \, \exp{\left( \frac12 \, \theta^T\, A \, \theta + \theta^T\, B\,  \eta\right)}.
\ee
Here $A$, $B$ are matrices of size $k\times k$ and $k\times m$ for some $k=O((n+m)^2)$.
 Matrix elements of $A$ and $B$ are linear functionals of $A_1,\ldots,A_n$
and $B_1,\ldots,B_n$. One can compute $A$ and $B$ in time $O(k)$.
Furthermore, if $G$ has bounded vertex degree then the same statement holds for $k=O(m+n)$.
\end{theorem}

Before going into technical details let us explain what is the main difficulty in representing the contracted tensor $T_V$
by a single Gaussian integral.
The point is that the convolution formula Eq.~(\ref{TuTv}) holds only if the edges incident to
the vertices $u,v$
are ordered in a consistent way as shown on Fig.~4.
If the orderings are not consistent, an extra sign may appear while commuting the variables
living on the contracted edges towards each other.
Assume one
wants to contract the combined vertex $u\star v$ with some third vertex $w$.
If the ordering of edges at the
combined vertex $u\star v$ is not consistent with the ordering at $w$,
one has to perform a cyclic shift of indexes in the tensor $T_{u\star v}$  and/or $T_w$ before
one can directly apply the formula Eq.~(\ref{TuTv}) to $T_{u\star v}$ and $T_{w}$.
Therefore, in general one can not represent the tensor $T_{u\star v\star w}$
obtained by contracting $u,v,w$ as a single Gaussian integral.

In order to avoid the problem with inconsistent edge orderings
we shall contract an open matchgate tensor network in two stages. At the first stage one
simulates each tensor $T_u$ by a matching sum of some planar graph
as explained in Section~\ref{subs:matchgate=gaussian}.
It yields an open tensor network in which every tensor
has a linear generating function (since every vertex must have exactly one incident edge).
At the second stage one represents the contraction of such a network by
a single convolution integral analogous to Eq.~(\ref{TuTv}). The problem with inconsistent
edge ordering will be addressed by choosing a proper orientation on every edge (which affects
the definition of monomials $\theta(e)$ in Eq.~(\ref{TuTv})). One can regard this approach as a generalization
of the original Kasteleyn's method~\cite{Kasteleyn61} to the case of a matching sum with "boundary conditions".

\begin{dfn}
A tensor $T$ is called linear if it has a linear generating function, $T=\sum_{a=1}^n w_a\, \theta_a$.
\end{dfn}
Clearly, any linear tensor $T$ can be mapped to $T(\theta)=\theta_1$ by a linear
change of variables. Lemma~\ref{lemma:matchgate2} implies that $T(\theta)=\theta_1$
is a matchgate. Therefore any linear tensor is a matchgate, see Lemma~\ref{lemma:inv}.
\begin{dfn}
Orientation of a graph $G=(V,E)$ is an antisymmetric matrix $A$ of size $|V|\times |V|$ such that
\[
A_{u,v}=\left\{ \ba{rcl} \pm 1 &\mbox{if} & (u,v)\in E,\\ 0 && \mbox{otherwise}.\\ \ea \right.
\]
An edge $(u,v)\in E$ is oriented from $u$ to $v$ iff $A_{u,v}=1$.
\end{dfn}
Recall that we represent each tensor $T_u$ by a generating function $T_u(\theta)$ that
depends on Grassmann variables $(\theta_{u,1},\ldots,\theta_{u,d(u)})$
associated with the edges incident to $u$, see Eq.~(\ref{T_u(theta)}).
Given an orientation $A$  of the graph $G$ and an edge $e=(u,v)\in E$ with the
labels $e_j^u\in E(u)$ and $e_k^v\in E(v)$, define
\be
\label{thetatheta}
\theta(e)=A_{u,v}\, \theta_{u,j}\, \theta_{v,k},
\quad
\int D\theta(e) = A_{u,v}\, \int d\theta_{v,k} \, \int \, d\theta_{u,j},
\quad \mbox{and} \quad
\int_{e\in E_{int}} D\theta(e)=\prod_{e\in E_{int}} \int D\theta(e).
\ee
Note that $\theta(e)$ and $\int D\theta(e)$ are symmetric under the transposition of $u$ and $v$.
\begin{lemma}
\label{lemma:open}
Let $T_V$ be a tensor obtained by contraction of an open tensor network on a graph
$G=(V,E)$.
Assume that all tensors in the network are linear. Then there exists an
orientation $A$ and an ordering of the vertices $V=\{v_1,v_2,\ldots,v_n\}$ such that
\be
\label{TuTv_open}
T_V=\int_{e\in E_{int}} D\theta(e)\;  T_{v_1} T_{v_2} \cdots T_{v_n}\, \exp{\left( \sum_{e\in E_{int}} \theta(e)\right)}.
\ee
The orientation and the ordering can be found in time $O(n)$.
\end{lemma}
\noindent
{\it Remark 1:} The generating function of $T_V$ is defined for the ordering of the
external edges in which they appear as one circumnavigates the boundary of the disk
anticlockwise.
The order of variables in $T_V$ corresponds to the counterclockwise order of the external edges.
\begin{proof}
Without loss of generality $G$ is a $2$-connected graph\footnote{If $G$ has a cut-vertex $u$ one can always
add an extra edge to some pair of nearest neighbors of $u$ in order to make $G$ $2$-connected.
The new edge must be assigned a zero weight in the two tensors it belongs to. Since the new edge
does not contribute to $T_V$ it can be safely removed at the end of the analysis.}.
Then the boundary of the outer face of $G$ is a closed loop without self-intersections. Let us denote it $\Gamma_{out}$.
Mark  some vertex in $\Gamma_{out}$ that has at least one incident external edge
(if there are no external edges, mark an arbitrary vertex).
Let $\Gamma_{out}=\{1,2,\ldots,m\}$
be an ordered list of all vertices on the outer face of $G$  corresponding to circumnavigating $\Gamma_{out}$
anticlockwise starting from the marked vertex. Extend the ordering of vertices to the rest of $V$ in an arbitrary way, so that
$V=\{1,2,\ldots,n\}$ and the first $m$ vertices belong to $\Gamma_{out}$.
\begin{dfn}
Let $G$ be a planar graph with the vertices ordered as described above.
A Kasteleyn orientation (KO) of $G$ is an orientation $A$ such that \\
(1) The number of c.c.w. oriented edges in the boundary of any face of $G$ is odd (except for the outer face).\\
(2) $A_{1,2}=A_{2,3}=\cdots=A_{m-1,m}=1$.
\end{dfn}
\noindent
{\it Remark:} The standard definition of a KO requires that (1) holds for all faces
of $G$ including the outer face and does not require (2), see for example~\cite{Reshetikhin06}.
By abuse of definitions we shall
apply the term KO to orientations satisfying (1),(2). The standard definition is not suitable
for our purposes because $G$ may have odd number of vertices
while the standard KO exists only on graphs with even number of vertices. The condition~(2)
is needed to ensure consistency between different "boundary conditions".
Example of a KO is shown on Fig.~6.
\begin{prop}
\label{prop:Kasteleyn}
Any  planar graph has a KO. It can be found in a linear time.
\end{prop}
We postpone the proof of the proposition until the end of the section. Let us choose
the orientation $A$ in Eq.~(\ref{thetatheta}) as a KO of the graph obtained from $G$
by removing all external edges. Let us verify that the contracted tensor $T_V$ can be
written as in
Eq.~(\ref{TuTv_open}).

 Indeed, let $S\subseteq E_{ext}$
be a subset of external edges such that any vertex in $\{1,\ldots,m\}$ has at most
one incident edge from $S$.
(Below we shall consider only such sets $S$ without explicitly mentioning it.)
Let $\partial S$ be a set of vertices that have an
incident edge from $S$ (clearly all such vertices belong to the outer face).
For any $S$ as above and any $\partial S$-imperfect matching $M\in \calM(G,\partial S)$ define
a subset of Grassmann variables
\[
\Omega(S,M)=\{ (u,j)\, : \,   u\in V, \quad \mbox{and} \quad e^u_j \in S\cup M\}.
\]
In other words, $(u,j)\in \Omega(S,M)$ iff $\theta_{u,j}$ is a Grassmann variable that live
on some edge of $S\cup M$. Note that there are two Grassmann variables living on any
internal edge and one variable living on any external edge. Thus for any $S$ and $M$
the set $\Omega(S,M)$ contains $n$ variables.
Define a normally  ordered monomial
\be
\label{order1}
\prod_{(u,j)\in \Omega(S,M)} \theta_{u,j}
\ee
as a product of all variables in $\Omega(S,M)$ ordered according to
\be
\label{corder}
(\theta_{1,1},\ldots,\theta_{1,d(1)},\theta_{2,1},\ldots,\theta_{2,d(2)},\ldots,\theta_{n,1},\ldots,\theta_{n,d(n)}).
\ee
Define also $M$-ordered monomial
\be
\label{order2}
\prod_{(u,j)\, : \, e^u_j\in S} \theta_{u,j}\, \prod_{e\in M} \theta(e),
\ee
where the order in the first product must agree with the chosen ordering of edges in $E_{ext}$,
see Fig.~6.
Clearly the two products Eqs.~(\ref{order1},\ref{order2}) coincide up to a sign that we shall denote
$\mathrm{sgn}(M)$. In order to prove Lemma~\ref{lemma:open} it suffices to show that
\be
\label{sgn(M)=1}
\mathrm{sgn}(M)=1 \quad \mbox{for all $\partial S$-imperfect matchings $M$, for all $S\subseteq E_{ext}$}.
\ee
Indeed, denoting  $T_u=\sum_{j=1}^{d(u)} w^u_j\, \theta_{u,j}$ one can rewrite Eq.~(\ref{TuTv_open}) as
\bea
T_V&=&\sum_{S\subseteq E_{ext}}
\sum_{M\in \calM(G,\partial S)}\, \int_{e\in E_{int}} \, D\theta(e)\,
\prod_{(u,j)\in \Omega(S,M)}w^u_j \theta_{u,j} \prod_{e\notin M} \theta(e)\nn \\
&=& \sum_{S\subseteq E_{ext}} \prod_{(u,j)\, : \, e^u_j\in S} \theta_{u,j}\,
\sum_{M\in \calM(G,\partial S)}\,
\mathrm{sgn}(M)\, \prod_{(u,j)\, : \, e^u_j\in M} w^u_j.
\eea
Assuming $\mathrm{sgn}(M)\equiv 1$ one can identify the sum over $M\in \calM(G,\partial S)$
with the component of the contracted tensor $T_V$ in which the subset $S$ of external edges
carries index $1$.

Note that for any $S\subseteq E_{ext}$ and  any $\partial S$-imperfect matching $M$
each vertex $u\in V$ contributes exactly one variable to $\Omega(S,M)$. Indeed, at every vertex $u\in V$
there is either one incident edge from $M$ or one incident external edge.
All other edges incident to $u$ and the variables living on these edges can be ignored
as far as computation of $\mathrm{sgn}(M)$ is concerned.
Therefore one can compute the
sign $\mathrm{sgn}(M)$ by introducing auxiliary Grassmann variables $\eta=(\eta_1,\ldots,\eta_n)$
associated with vertices of $G$ and comparing the
normal ordering of $\eta$
( the one in which the indexes increase from the left to the
right)
with the $M$-ordering of $\eta$, namely
\[
\prod_{u\in \partial S} \eta_u\, \prod_{e\in M} \eta(e) = \mathrm{sgn}(M)\, \eta_1\eta_2 \cdots \eta_n,
\quad \mbox{where} \quad \eta(e)=A_{u,v}\, \eta_u \eta_v \quad \mbox{if} \quad e=(u,v).
\]
Here the ordering in the first product is normal
while the ordering in the second product may be
arbitrary since $\eta(e)$ is a central element.
Consider any subsets $S,S'\subseteq E_{ext}$.
Given any $\partial S$-imperfect matching $M$ and $\partial S'$-imperfect matching $M'$ define
a relative sign
\be
\label{rel_sign}
\mathrm{sgn}(M,M')\eqdef \mathrm{sgn}(M)\, \mathrm{sgn}(M'),
\ee
such that
\be
\label{rel_sign1}
\prod_{u\in \partial S} \eta_u\, \prod_{e\in M} \eta(e) =\mathrm{sgn}(M,M')\,
\prod_{u\in \partial S'} \eta_u\, \prod_{e\in M'} \eta(e).
\ee
In order to compute  $\mathrm{sgn}(M,M')$ consider the symmetric difference $M\oplus M'$.
It consists of a disjoint union of even-length cycles $C_1,\ldots,C_p$ and open paths $\Gamma_1,\ldots,\Gamma_q$
such that every path $\Gamma_j$ has both its endpoints in the symmetric difference $\partial S\oplus \partial S'$.
Given a path $\Gamma_j$ with endpoints $s,t\in \partial S\oplus \partial S'$, $s<t$ let us orient $\Gamma_j$ from $s$
to $t$. Now one can compute the relative sign  as follows.
\begin{prop}
\label{prop:relative_sign}
Consider any subsets $S,S'\subseteq E_{ext}$.
Let $C_1,\ldots,C_p$ and $\Gamma_1,\ldots,\Gamma_q$ be the cycles and the paths formed by
$M\oplus M'$ for
some $\partial S$-imperfect matching $M$ and some $\partial S'$-imperfect matching $M'$.
For a path $\Gamma_j$ connecting vertices $s,t\in \partial S\oplus  \partial S'$ on the outer face such that $s<t$
let $\omega(\Gamma_j)=1$ if the interval $(s,t)$ contains odd number  of vertices from $\partial S$
and $\omega(\Gamma_j)=0$ if this number is even. Then
\be
\label{sigma(M,M')}
\mathrm{sgn}(M,M')=(-1)^p\, \prod_{j=1}^p \Phi(C_j) \; \prod_{k=1}^q (-1)^{\omega(\Gamma_k)}\,  \Phi(\Gamma_k),
\ee
where
\[
\Phi(C_j)=\prod_{(u,v)\in C_j} A_{u,v} \quad \mbox{and} \quad \Phi(\Gamma_k)=\prod_{(u,v)\in \Gamma_k} A_{u,v}.
\]
 \end{prop}
 \noindent
 {\it Remark 1:}  The definition of $\omega(\Gamma_j)$ is symmetric under exchange of $S$ and $S'$. Indeed,
the overall number of  vertices from  $\partial S\oplus \partial S'$ contained in the interval $(s,t)$
 is even since these vertices are pairwise connected by $\Gamma$'s. The remaining vertices
 of $(s,t)$ either belong to both sets $S,S'$ or belong to neither of them. \\
{\it Remark 2:} The product $\prod_{(u,v)\in \Gamma_k} A_{u,v}$ gives the parity of the number of edges
in $\Gamma_k$ whose orientation determined by $A$ disagrees with the chosen orientation of $\Gamma_k$.
The product  $\Phi(C_j)$ does not depend on how one chooses orientation of $C_j$
since  every cycle $C_j$ has even length.
\begin{proof}
Indeed, one can easily check that for every cycle $C_j$  one has
\be
\label{flux3prop}
\prod_{e\in C_j\cap M} \eta(e)=-\Phi(C_j)\, \prod_{e\in C_j \cap M'} \eta(e).
\ee
Therefore changing the $M$-ordering to the $M'$-ordering in a cycle $C_j$ contributes a factor
$-\Phi(C_j)$ to the relative sign $\mathrm{sgn}(M,M')$.
Consider now a path $\Gamma_j$ connecting vertices $s,t\in \partial S\oplus \partial S'$ where $s<t$.
Let us argue  that changing the $M$-ordering to the $M'$-ordering on the path $\Gamma_j$
contributes a factor $(-1)^{\omega(\Gamma_j)}\, \Phi(\Gamma_j)$ to the relative sign $\mathrm{sgn}(M,M')$.
Indeed, one can easily check the following identities:
\bea
s,t\in S &:& \eta_s\eta_t \prod_{e\in \Gamma_j\cap M} \eta(e) = \Phi(\Gamma_j)\, \prod_{e\in \Gamma_j\cap M'} \eta(e),\nn \\
s,t\in S' &:& \mbox{the same as above up to $M\leftrightarrow M'$}, \nn \\
s\in S, t\in S' &:& \eta_s \prod_{e\in \Gamma_j\cap M} \eta(e) =\Phi(\Gamma_j)\, \eta_t \, \prod_{e\in \Gamma_j\cap M'} \eta(e),\nn \\
s\in S', t\in S &:&\mbox{the same as above up to $M\leftrightarrow M'$}. \nn
\eea
Consider as example the case $s,t\in S$. Bringing the variables $\eta_s$ and $\eta_t$ together in
the monomial $\prod_{u\in \partial S} \eta_u$ introduces  an extra  sign $(-1)^{\omega(\Gamma_j)}$.
Taking into account that $\eta(e)$ are central elements and using the first identity above one concludes that
\[
\prod_{u\in \partial S} \eta_u \, \prod_{e\in \Gamma_j\cap M}\eta(e) = (-1)^{\omega(\Gamma_j)}\, \Phi(\Gamma_j)\,
\prod_{u\in \partial S\backslash \{s,t\}} \eta_u \prod_{e\in \Gamma_j\cap M'} \eta(e).
\]
Other three cases can be considered analogously using Remark~1 above.
Combing it with Eq.~(\ref{flux3prop}) one arrives to
Eq.~(\ref{sigma(M,M')}).
\end{proof}
Let us  proceed  with the proof of Lemma~\ref{lemma:open}.
The first condition in the definition of KO implies\footnote{This is the well-known property of a Kasteleyn orientation
which we prove below for the sake of completeness.}
 that $\Phi(C_j)=-1$ for all cycles $C_j$.
Indeed, consider any particular cycle $C_j$ and let $N_0,N_1,N_2$ be the number
of vertices, edges, and faces in the subgraph bounded by $C_j$.
The Euler formula implies that $N_0+N_2-N_1=1$.
Denote also $N_1^{int}$ the number of {\it internal } edges, i.e., edges having at least one endpoint in the
interior of $C_j$.  Since $C_j$ has even length, $N_1^{int}$ has the same parity as $N_1$.
Furthermore, since all vertices of the subgraph bounded by $C_j$ are paired by $M$ (and by $M'$), $N_0$ is even.
Since $\Phi(C_j)$ can be regarded as a parity of c.c.w. oriented edges in $C_j$
and each internal edge is c.c.w. oriented with respect to one of the adjacent faces
the property~(1) of KO yields
\be
\label{flux1}
\Phi(C_j)=(-1)^{N_2+N_1^{int}}=(-1)^{N_2+N_1}=(-1)^{1+N_0}=-1.
\ee
Therefore Proposition~\ref{prop:relative_sign} implies
\be
\label{flux1'}
\mathrm{sgn}(M,M') =\prod_{k=1}^q (-1)^{\omega(\Gamma_k)}\, \Phi(\Gamma_k).
\ee
Let us now show that
\be
\label{flux2}
(-1)^{\omega(\Gamma_k)}\, \Phi(\Gamma_k)=1
\ee
for all paths $\Gamma_k$. Indeed, let $s,t\in S\oplus S'$ be the
starting and the ending vertices of $\Gamma_k$. Consider a
path $\Gamma_k^*$  obtained by passing from $t$ to $s$ along
the boundary of the outer face $\Gamma_{out}$ in the clockwise direction.
Let $N_0,N_1,N_2$ be the number
of vertices, edges, and faces in the subgraph bounded by a cycle $\Gamma_k \cup \Gamma_k^*$.
Denote also $N_1^{int}$ the number of edges that have at least one endpoint in the interior of
$\Gamma_k \cup \Gamma_k^*$.
The Euler formula implies that $N_0+N_2-N_1=1$.
Note that $\Phi(\Gamma_k)$ can be regarded as the parity of the number of edges
in $\Gamma_k$ whose orientation determined by $A$ corresponds to c.c.w. orientation
of the cycle $\Gamma_k \cup \Gamma_k^*$.
Repeating the  arguments leading to Eq.~(\ref{flux1}) and noting that
all edges of the cycle $\Gamma_k \cup \Gamma_k^*$ belonging to $\Gamma_k^*$ are oriented c.c.w.
one gets
\be
\label{flux3}
\Phi(\Gamma_k)=(-1)^{|\Gamma_k^*|+N_2+N_1^{int}}=(-1)^{|\Gamma_k|+N_2+N_1}=(-1)^{|\Gamma_k|+N_0+1}.
\ee
Here $|\Gamma_k|$ and $|\Gamma_k^*|$
are the numbers of edges in the two paths. Consider three possibility:\\
{\it Case~1:} $s,t\in \partial S$.  Then
$|\Gamma_k|$ is odd and thus $\Phi(\Gamma_k)=(-1)^{N_0}$.
All $N_0$ vertices of the graph bounded by $\Gamma_k \cup \Gamma_k^*$
are paired by the matching $M$ except for $s,t$ and those belonging to $\partial S$
and lying on the interval
$(s,t)$. Therefore the parity of $N_0$ coincides with $\omega(\Gamma_k)$ and we arrive to Eq.~(\ref{flux2}).\\
{\it Case~2:} $s,t\in \partial S'$. The same as Case~1  (see Remark~1 after Proposition~\ref{prop:relative_sign}).\\
{\it Case~3:} $s\in \partial S$, $t\in \partial S'$ (or vice verse).Then $|\Gamma_k|$ is even and thus
$\Phi(\Gamma_k)=(-1)^{N_0+1}$.
All $N_0$ vertices of the graph bounded by $\Gamma_k \cup \Gamma_k^*$
are paired by the matching $M$ except for $s$ (or except for $t$)  and those belonging to $\partial S$
and lying on the interval
$(s,t)$. Therefore the parity of $N_0$ coincides with $\omega(\Gamma_k)+1$ and we arrive to  Eq.~(\ref{flux2}).

Combining Eqs.~(\ref{flux1},\ref{flux2}) and Proposition~\ref{prop:relative_sign}
we conclude that $\mathrm{sgn}(M,M')=1$ for all $M$ and $M'$. Thus either
$\mathrm{sgn}(M)=1$ for all $M$ or $\mathrm{sgn}(M)=-1$ for all $M$.
One can always exclude the latter possibility by applying a {\it gauge transformation} to
the orientation $A$. A gauge transformation at a vertex $u\in V$ reverses
orientation of all edges incident to $u$.
Let us say that a vertex $u\in V$ is {\it internal} if does not belong to the outer face of $G$.
 Clearly a gauge transformation
at any internal  vertex $u$ maps a KO to a KO and flips the sign $\mathrm{sgn}(M)$ for all
$M$. Thus it suffices to consider the case when  $G$ does not have internal vertices (i.e. $G$ is an outerplanar graph).
If $m=n$ is even, a matching  $M=\{(1,2),(3,4),\ldots,(m-1,m)\}$ has sign $\mathrm{sgn}(M)=1$
due to property (1) of a KO and thus all matchings have sign $+1$.
If $m=n$ is odd one can apply the same argument using  a matching $M=\{(2,3),(4,5),\ldots,(m-1,m)\}$
(recall that the vertex $1$ has at least one external edge and thus it can be omitted in $M$).
\end{proof}

\begin{proof}[Proof of Theorem~\ref{thm:partial_contraction}]
Let $n_e$ be the number of internal edges in the graph $G$, so that $|E|=m+n_e$.
Since $G$ is a planar graph, $n_e=O(n)$, see for example~\cite{Diestel},
and thus $|E|=O(n+m)$.
Denote degree of a vertex $u\in V$ by $d(u)$ (it includes both internal and external edges).
Applying Theorem~\ref{thm:cycles} one can simulate the tensor $T_u$ at any vertex $u\in V$ by
a matching sum of some planar  graph $G_u$ with $O(d(u)^2)$ vertices.
Combining the graphs $G_u$ together one gets an open tensor network $G'=(V',E')$
in which all tensors are linear. The network $G'$ has
$m$ external edges.
The number of vertices $n'$ in the network $G'$ can be
bounded as
$n'=\sum_{u\in V} O(d(u)^2) = O((\sum_{u\in V} d(u))^2) =O(|E|^2)=O((m+n)^2)$.
If $G$ has bounded degree one gets $n'=\sum_{u\in V} O(d(u)^2)=O(n)$.
Thus in both cases $n'=O(k)$, where $k$ is defined in the statement of the theorem.
It follows from Theorem~\ref{thm:cycles}
that the edge weights in the matching sums are linear functions of the matrix elements of $A_1,\ldots,A_n$
and $B_1,\ldots,B_n$. Let $n_e'$ be the number of internal edges in $G'$.
Since $G'$ is a planar graph, $n_e'=O(n')=O(k)$. Thus the total number of edges in $G'$
is $|E'|=n_e'+m=O(k)$.
Invoking Lemma~\ref{lemma:open} we need to introduce
a pair of Grassmann variables for every internal edge of $G'$ and one variable
for every external edge. Thus the total number of Grassmann variables is $O(k)$.
It determines the number of variables in  the vector $\theta$ in Eq.~(\ref{integral_main}).
Representing linear tensors $T_j$ as  Gaussian integrals, namely
\[
T_j = \int d\mu \exp{(\mu\, T_j)},
\]
one can combine the multiple integrals in Eq.~(\ref{TuTv_open})
into a single Gaussian integral Eq.~(\ref{integral_main}) with the matrix $A$
having a dimension $O(k)\times O(k)$ and $B$ having a dimension $O(k)\times m$.
Thus $A$ and $B$ have  the  desired properties.
\end{proof}

\begin{proof}[Proof of Proposition~\ref{prop:Kasteleyn}]
Let $G=(V,E)$ be a planar graph with $n$ vertices such that the outer face of $G$ is a simple loop.
An orientation $A$ satisfying (1) can be constructed using the algorithm of~\cite{Reshetikhin06}.
For the sake of completeness we  outline it below. Let $G^*=(V^*,E)$ be the dual graph
such that  each face of $G$ contributes one vertex to $G^*$ (including the outer face).
Let $T$ be a spanning tree of $G^*$ such that the root of $T$ is the outer face of $G$.
One can find $T$ in time $O(|V|+|E|)=O(n)$ since for planar graphs $|E|=O(|V|)$.
Assign an arbitrary orientation to those edges of $G$ that do not belong to $T$.
By moving from the leaves of $T$ to the root assign the orientation to all edges of $T$.
Note that for every vertex $u$ of $T$ which is not the root
 the orientation of an edge $e$ connecting $u$ to its ancestor is uniquely determined by (1).
We obtained an orientation of all edges of $G$ satisfying (1).

In order to satisfy (2) one can apply a series of {\it gauge transformations}. A gauge transformation
at a vertex $u\in V$ reverses orientation of all edges incident to $u$. Clearly it preserves
the property (1). Applying if necessary a gauge transformation at the vertices $\{1,2,\ldots,m-1\}$ one can
satisfy~(2).
 \end{proof}

\subsection{Contraction of matchgate networks with a single vertex}
\label{subs:1vertex}
In this section we explain how to contract a matchgate tensor network $\calT$
that consists of a single vertex $u$ with $m$ self-loops
embedded into a surface  $\Sigma$ of genus $g$ without self-intersections.
Example of such a network with $m=3$ and $g=1$ is shown on Fig.~7.
Let $T$ be a tensor of rank $2m$ associated with $u$.
Clearly the contraction value $c(\calT)$
depends only on the pairing pattern indicating what indexes of $T$ are contracted with each other.
It will be convenient to represent the pairing pattern by a {\it pairing graph} $P=(V,E)$ with  a set
of vertices $V=\{1,2,\ldots,2m\}$ such that 
a pair of vertices $(a,b)$ is connected by an edge
iff the indexes $a,b$ of the tensor $T$ are
contracted with each other (connected by a self-loop). 
By definition $P$ consists of $m$ disjoint edges.
Let us embed $P$ into a disk such that
all the vertices of $P$ lie on the boundary of the disk and their order corresponds to
circumnavigating  the boundary anticlockwise. The edges of $P$ are represented by
arcs lying inside the disk, see Fig.~7.  One can always draw the arcs such that 
there are only pairwise intersection points.

Introduce an auxiliary tensor $R$ of rank $2m$ such that
\[
R(x)=\left\{ \ba{rcl}
1 &\mbox{if} & x_a=x_b \quad \mbox{for all $(a,b)\in E$}, \\
0 & \mbox{if} & x_a\ne x_b \; \mbox{for some $(a,b)\in E$}.\\
\ea\right.
\]

\begin{figure}
\label{fig:pairing}
\centerline{
\mbox{
 \includegraphics[height=3cm]{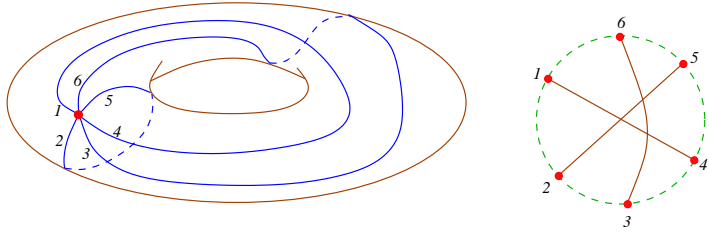}
 }}
\caption{Left: a tensor network with a single vertex embedded into a torus. Right: the pairing graph $P$. }
\end{figure}

The contraction value of $\calT$ can be represented as
\be
\label{TR}
c(\calT)=\sum_{x\in \{0,1\}^{2m}}  T(x)\, R(x).
\ee
Let $\theta=(\theta_1,\ldots,\theta_{2m})$ and $\eta=(\eta_1,\ldots,\eta_{2m})$ be Grassmann variables
and $T(\theta)$, $R(\eta)$ be  the generating functions of $T$ and $R$. 
\begin{prop}
Let $\epsilon(T)=0,1 $ for even  and odd tensors $T$ respectively . Then
\be
\label{TR1}
c(\calT)=i^{\epsilon(T)}\,
\int D\theta \int D\eta \, T(\theta)\, R(\eta)\, \exp{(i\, \theta^T\eta)}.
\ee
\end{prop}
\begin{proof}
A non-zero contribution to the integral comes from the 
terms in which $T(\theta)$ contributes monomial $T(x)\, \theta(x)$
and $R(\eta)$ contributes monomial $R(x)\, \eta(x)$ for some $x\in \{0,1\}^{2m}$. 
A simple algebra shows that 
for any $x\in \{0,1\}^{2m}$ one has the following identity
\[
\theta(x)\, \eta(x)\, \prod_{a\, : \, x_a=0} i \theta_a \eta_a  =
i^{-|x|}\, (-1)^{|x|\, (|x|-1)/2}\, \theta(1^{2m})\, \eta(1^{2m}),
\]
where $|x|$ is the Hamming weight of $x$. 
Taking into account that $T(x)=0$ unless $|x|$ has parity $\epsilon(T)$
one gets
\[
i^{-|x|}\, (-1)^{|x|\, (|x|-1)/2}=i^{-\epsilon(T)}.
\]
Since $\int \, D\theta \int \, D\eta\,  \theta(1^{2m})\, \eta(1^{2m})=1$, one gets Eq.~(\ref{TR1}).
\end{proof}
In general $R$ is not a matchgate tensor because the chosen planar embedding of the 
pairing graph may have edge crossing points. For example, assume that $P$ has $4$ vertices
$\{1,2,3,4\}$ and two edges $(1,3)$, $(2,4)$ (which can be realized on a torus). 
Then the non-zero components of $R$
are $R(0000)=R(1010)=R(0101)=R(1111)=1$. Substituting them into the matchgate identities
Eq.~(\ref{Grank4})
for even rank-$4$ tensors one concludes that $R$ is not a matchgate.

Let us order the edges of $P$ in an arbitrary way, say, $E=\{e_1,e_2,\ldots,e_m\}$.
For any edges $e_p,e_q\in E$ let
$N_{p,q}$ be the  the number of self-intersections of $e_p,e_q$ in the
planar embedding shown on Fig.~7. Since we assumed that all intersections are pairwise,
 $N_{p,q}$ takes only values $0,1$, i.e., $N$ is a symmetric binary matrix. Let us also agree that $N_{p,p}=0$.
We shall see later that the tensor $R$ can be represented as a linear combination of $2^r$
matchgate tensors, where $r$ is a binary rank of the matrix $N$. It is crucial that the rank of $N$
can be bounded by  the genus $g$ of the surface $\Sigma$.
\begin{lemma}
\label{lemma:rank}
The matrix $N$ has binary rank at most $2g$.
\end{lemma}
\begin{proof}
Let us cut a small disk $D$  centered at the vertex $u$ from the surface $\Sigma$,
embed the pairing graph $P$ into the disk $D$ as shown on Fig.~7
and glue the disk back to the surface $\Sigma$. Thus given any 
self-loop $\alpha$ connecting indexes $a$ and $b$ of the tensor $T$,
a small section  of $\alpha$ lying inside $D$ is replaced by 
an edge $e=(a,b)\in E$ of the pairing graph.
 We get a family of $m$ closed loops embedded into
$\Sigma$. The loops may have pairwise intersection points inside the disk $D$.
Let $\alpha_p$ be a loop that contains an edge $e_p\in E$.
To every loop $\alpha_p$ one can assign its homological class $[\alpha_p]\in H_1(\Sigma,\ZZ_2)$
in the first homological group of $\Sigma$ with binary coefficients. Since all
intersection points between the loops  are contained in the disk $D$, we get
\[
N_{p,q}=\omega([\alpha_p],[\alpha_q]),
\]
where $\omega\, : \, H_1(\Sigma,\ZZ_2)\times H_1(\Sigma,\ZZ_2) \to \{0,1\}$ is
the  intersection form. It is well known that
the intersection form defined on a surface $\Sigma$ of genus $g$ has rank $2g$. Therefore, $N$ has rank at most $2g$.
\end{proof}
Given any edge $e\in E$, let  $l(e),r(e)\in V$ be the two endpoints of $e$ such that $l(e)<r(e)$.
Denote $\eta(e)=\eta_{l(e)}\, \eta_{r(e)}$. The generating function for the tensor $R$ can be written as
\be
\label{R(eta)}
R(\eta) =
\sum_{y\in \{0,1\}^m}
 (-1)^{\frac12\, y^T\, N \, y} \, \prod_{e\in y} \eta(e), \quad \mbox{where} \quad
\eta(e)=\eta_{l(e)}\, \eta_{r(e)}.
\ee
Here we identified a binary string $y\in \{0,1\}^m$ with the subset of edges
$e_a\in E$ such that $y_a=1$.
Indeed, for any $x\in \{0,1\}^{2m}$ such that $R(x)=1$
one  has to regroup the factors in $\eta(x)$ to bring together
variables corresponding to the same edge. It yields an extra minus sign
for every pair of intersecting edges in $y$.
Since every pair of edges $e_a,e_b$ contributes a sign $(-1)^{N_{a,b} \, y_a y_b}$,
we arrive to Eq.~(\ref{R(eta)}).

Consider binary Fourier transform of the function $(-1)^{\frac12\, y^T\, N \, y}$,
\be
\label{Fourier}
f(z)\eqdef\frac1{2^m} \sum_{y\in \{0,1\}^m} (-1)^{\frac12\, y^T\, N \, y + z\cdot y}, \quad z\in \{0,1\}^m.
\ee
Clearly $f(z)=0$ unless $z\in \mathrm{Ker}(N)^\perp$, where $\mathrm{Ker}(N)=\{ y\in \{0,1\}^m \, : \, Ny=0\}$
is the zero subspace of $N$. If $N$ has rank $r$, the zero subspace of $N$ has dimension
$m-r$ and thus $\mathrm{Ker}(N)^\perp$ has dimension $r$. Let us order all the vectors
of $\mathrm{Ker}(N)^\perp$ in an arbitrary way
\[
\mathrm{Ker}(N)^\perp=\{z^1,\ldots,z^{2^r}\}.
\]
Applying the reverse Fourier transform one gets
\be
\label{short_sum}
(-1)^{\frac12\, y^T\, N \, y}= \sum_{a=1}^{2^r} f(z^a)\, (-1)^{y\cdot z^a}.
\ee
By Lemma~\ref{lemma:rank} the number of terms in the sum above is bounded by $2^{2g}$.
Substituting Eq.~(\ref{short_sum}) into Eq.~(\ref{R(eta)}) we arrive to
\be
\label{GaussianR}
R(\eta)= \sum_{a=1}^{2^r} f(z^a)\, \exp{\left( \sum_{e\in E} (-1)^{(z^a)_e} \, \eta(e) \right)},
\ee
where $(z^a)_e$ is the component of the vector $z^a$ corresponding to an edge $e$.
It shows that $R$ is indeed a linear combination of $2^r$ matchgate tensors with $r\le 2g$.

In order to get an explicit formula for  the contraction value Eq.~(\ref{TR}) let us introduce an auxiliary
$2m\times 2m$ matrix
 \[
A_{j,k}=\left\{ \ba{rcl}
+1 &\mbox{if} & \mbox{$j=l(e)$, $k=r(e)$ for some $e\in E$,}\\
-1 &\mbox{if} &\mbox{$j=r(e)$, $k=l(e)$ for some $e\in E$}\\
0 &&\mbox{otherwise}\\
\ea
\right.
\]
Introduce also auxiliary diagonal $2m\times 2m$ matrices $D^a$, $a=1,\ldots,2^r$ such that
\[
(D^a)_{j,j}= \left\{
\ba{rcl}
(-1)^{(z^a)_e} &\mbox{if}& \mbox{$j=l(e)$ for some $e\in E$},\\
1 &&\mbox{otherwise}.\\
\ea\right.
\]
Then Eq.~(\ref{GaussianR}) can be rewritten as
\be
R(\eta)= \sum_{a=1}^{2^r} f(z^a)\,
\exp{\left( \frac12\, \eta^T\, D^a\, A\, D^a\,  \eta\right)}.
\ee
Theorem~\ref{thm:canonical} implies that $T$ can be described by a generating function
\[
T(\theta)=C\exp{\left( \frac12\, \theta^T\, F \, \theta \right)}\int D\mu\, \exp{\left( \mu^T \, G\, \theta \right)},
\]
where $F$ and $G$ have size $2m\times 2m$ and $k\times 2m$ for some even integer $0\le k\le 2m$.
Using Eq.~(\ref{TR1}) one can express the contraction value $c(\calT)$ as a linear
combination of $2^r$ Gaussian integrals
\be
c(\calT)=C\, \sum_{a=1}^{2^r}
f(z^a)\,
\int D\theta\,  D\eta\, D\mu \, \exp{\left(
\frac12 \, \theta^T\, F\, \theta + \frac12 \, \eta^T \, D^a \, A \, D^a \, \eta + \mu^T\, G\, \theta  + i\, \theta^T\, \eta \right)}.
\ee
Introducing a $(4m+k)\times (4m+k)$ matrix
\[
M^a=\left[ \ba{ccc}
F & iI & -G^T \\
-iI & -D^a\, A\, D^a & 0 \\
G & 0 & 0 \\
\ea
\right]
\]
one finally gets
\be
\label{cv=5}
c(\calT)=C\, \sum_{a=1}^{2^r}
f(z^a)\, \pf{(M^a)}.
\ee
Computing $\pf{(M^a)}$ requires time  $O(m^3)$.  Lemma~\ref{lemma:rank} implies that the number of
terms in the sum is at most $2^{2g}$. 
Finally, as we show below one can compute $f(z^a)$ in time $O(m^3)$. 
Thus $c(\calT)$ can be computed in time $O(m^3)\, 2^{2g}$.

\begin{prop}
The function $f(z)$ in Eq.~(\ref{Fourier}) can be represented as
\be
\label{f(z)}
f(z)=\frac1{2^{r/2}}\, (-1)^{\frac12\, z^T\, M\, z}
\ee
for some matrix $M$ computable in time $O(m^3)$.
\end{prop}
\begin{proof}
Using Gaussian elimination any symmetric binary matrix $N$ with zero
diagonal  can be represented as $N=U^T\, \tilde{N}\, U$, where $U$ is a binary invertible matrix
and $\tilde{N}$ is a block diagonal matrix with $2\times 2$ blocks, 
\[
\tilde{N}=\bigoplus_{j=1}^{r/2} \left( \ba{cc} 0 &1 \\ 1 & 0 \\ \ea\right).
\] 
In particular, the rank of $N$ is always even.
The matrix $U$ can be found in time $O(m^3)$.
Performing a change of variable $y\to Uy$ in Eq.~(\ref{Fourier}) one gets
\be
\label{Fourier1}
f(z)=\frac1{2^m} \sum_{y\in \{0,1\}^m} (-1)^{\sum_{j=1}^{r/2} y_{2j-1} y_{2j} + \tilde{z}\cdot y}, \quad \tilde{z}=(U^{-1})^T\,z.
\ee
It follows that $f(z)=0$ unless $\tilde{z}_{r+1}=\ldots=\tilde{z}_m=0$.
Using an identity
\[
(-1)^{x_1\cdot x_2} =\frac12 \sum_{y_1,y_2=0,1} (-1)^{y_1\cdot y_2 + y_1 \cdot x_1 + y_2 \cdot x_2}
\]
one can rewrite Eq.~(\ref{Fourier1}) as
\[
f(z)=\frac{1}{2^{r/2}} (-1)^{\sum_{j=1}^{r/2}\, \tilde{z}_{2j-1}\, \tilde{z}_{2j}}= 
\frac{1}{2^{r/2}} (-1)^{\frac12\, z^T\, U^{-1} \, \tilde{N}\, (U^{-1})^T\, z}.
\]
We get the desired expression Eq.~(\ref{f(z)}) with $M=U^{-1} \, \tilde{N}\, (U^{-1})^T$.
\end{proof}

\subsection{The main theorem}
Theorem~\ref{thm:main} can be obtained straightforwardly from Theorem~\ref{thm:partial_contraction}
and the contraction algorithm for a network with a single vertex, see Section~\ref{subs:1vertex}. Indeed, let $M$ be a planar cut of $G$
with $m$ edges and $G_M$ be a subgraph obtained from $G$ by removing all edges of $M$.
By definition $G_M$ is contained in some region $D$ with topology of a disk.
Without loss of generality $D$ contains no edges from $M$ (otherwise one can 
remove these edges from $M$ getting a planar cut
with a smaller number of edges). Thus one can regard $G_M$ as an open tensor network
with   $2m$ external edges. Since $G_M$ contains all vertices of $G$, 
the network obtained by contraction of $G_M$ consists of a single vertex and $m$ self-loops.
As explained in the previous section, one can compute the contraction value of such
a network in time $O(m^3)\, 2^{2g}$.

In order to 
contract  $G_M$
one has to compute the Gaussian integral Eq.~(\ref{integral_main}). 
Theorem~\ref{thm:partial_contraction} guarantees that 
this integral involves matrices of size $k$, where
$k=O((n+m)^2)$ or $k=O(n+m)$ depending on whether the graph $G$
has bounded vertex degree. 
As explained in Section~\ref{subs:Gintegral}
the Gaussian integral with matrices of size $k$ can be computed in time $O(k^3)$.
Combining the two parts together one gets Theorem~\ref{thm:main}.

\section*{Acknowledgements}
The author acknowledge support by DTO through ARO contract number
W911NF-04-C-0098.

\section*{Appendix A}

Suppose $T_u$ and $T_v$ are matchgate tensors specified by their canonical generating functions
as in Eq.~(\ref{canonical}), that is
\[
T_\alpha = C_\alpha \, \exp{\left( \frac12\, \theta^T_\alpha \, A_\alpha \,
 \theta_\alpha \right)}\int D\mu_\alpha \, \exp{\left( \mu^T_\alpha \, B_\alpha\, \theta_\alpha \right)},
 \quad \mbox{where} \quad \alpha=u,v.
\]
Here $\theta_u=(\theta_{u,1},\ldots,\theta_{u,d(u)})$ and
$\theta_v=(\theta_{v,1},\ldots,\theta_{v,d(v)})$ are the two sets of Grassmann variables
associated with the vertices $u$ and $v$.
Denote also $\epsilon(T)$ the parity of a matchgate tensor $T$, that is, $\epsilon(T)=0$ ($\epsilon(T)=1$)
for even (odd) tensor $T$.
In the remainder of this section we explain how to express the canonical generating function
for the contracted tensor $T_{u\star v}$, see Eqs.~(\ref{Tu*v_def},\ref{TuTv}),
 in terms of the matrices $A_\alpha$, $B_\alpha$.

Applying Eq.~(\ref{TuTv}) one gets
\be
\label{misc1}
T_{u\star v}=C_uC_v\, \int_{e\in E(u,v)} \, D\theta(e) \,
\int D\mu_u \, \int D\mu_v \,
\exp{[ f(\theta_u,\theta_v,\mu_u,\mu_v)]},
\ee
where
\[
 f(\theta_u,\theta_v,\mu_u,\mu_v)=
 \sum_{\alpha=u,v} \, \frac12 \, \theta_\alpha^T A_\alpha\, \theta_\alpha + \mu_\alpha^T\, B_\alpha\, \theta_\alpha +
 \sum_{e\in E(u,v)} \theta(e).
\]
Let us split the vectors of Grassmann variables $\theta_u$, $\theta_v$ into
external and internal parts,
\[
\theta_u=(\theta_u^e,\theta_u^i) \quad  \mbox{and}  \quad \theta_v=(\theta_v^i,\theta_v^e),
\]
such that all internal variables are integrated out in $T_{u\star v}$.
Then one can rewrite the expression Eq.~(\ref{misc1})
as a product of a Gaussian exponent and the standard Gaussian integral
$I(K,L)$, see Eqs.~(\ref{GI2},\ref{GI3}),  for some matrices $K,L$ defined below,
\be
\label{misc2}
T_{u\star v}(\tau)=C_u C_v\,(-1)^{\frac{b(b-1)}2 + \epsilon(T_u) \epsilon(T_v)}\,  \exp{\left( \frac12\, \tau^T\, H \, \tau\right)}
\,
\int D\eta \exp{\left( \frac12 \, \eta^T\, K\, \eta + \eta^T\, L \, \tau\right)}.
\ee
Here we introduced auxiliary vectors of Grassmann variables
$\tau=(\theta_u^e,\theta_v^e)$, $\eta=(\theta_u^i,\theta_v^i,\mu_u,\mu_v)$.
The matrices $H,K,L$ above will be defined using a partition of matrices $A_\alpha$, $B_\alpha$
into "internal" and "external" blocks as follows:
\[
A_u=\left[ \ba{cc}  A_u^{ee} & A_u^{ei} \\ A_u^{ie} & A_u^{ii} \\ \ea \right],
\quad
A_v=\left[ \ba{cc}  A_v^{ii} & A_v^{ie} \\ A_v^{ei} & A_v^{ee} \\ \ea \right],
\quad
B_u=\left[ \ba{cc} B_u^e & B_u^i \\ \ea \right],
\quad
B_v=\left[ \ba{cc} B_v^i & B_v^e \\ \ea \right].
\]
Introduce also a square matrix $\bar{I}$ that has ones on the diagonal perpendicular to the main diagonal
and zeroes everywhere else. Then the matrices $H,K,L$ in Eq.~(\ref{misc2}) are defined as
\[
H=\left[ \ba{cc} A_u^{ee} & 0 \\ 0& A_v^{ee} \\ \ea \right],
\quad
K=\left[ \ba{cccc}
A_u^{ii} & \bar{I} & -(B_u^i)^T & 0 \\
& A_v^{ii} & 0 & -(B_v^i)^T \\
& & 0 & 0 \\
& & & 0\\
\ea\right],
\quad
L=\left[ \ba{cc}
A_u^{ie} & 0 \\
0 & A_v^{ie} \\
B_u^e & 0\\
0 & B_v^e \\
\ea
\right].
\]
Finally, the extra sign in Eq.~(\ref{misc2}) takes into account the difference between
the order of integrations in Eqs.~(\ref{misc1},\ref{misc2}). Summarizing, Eq.~(\ref{misc2})
together with the Gaussian integration formulas Eqs.~(\ref{GI2},\ref{GI3}) allow one
to write down the canonical generating function for the contracted tensor $T_{u\star v}$.

\end{document}